\documentclass[journal,onecolumn]{IEEEtran}
\usepackage{amsmath}
\usepackage{booktabs}
\usepackage{latexsym}
\usepackage{mathrsfs}
\usepackage{amsthm}
\usepackage{amssymb}
\usepackage{amsfonts}
\usepackage{amsbsy}

\usepackage[shortlabels]{enumitem}
\usepackage{url}
\usepackage{array}
\usepackage{pdflscape}
\usepackage{xcolor}
\usepackage{stmaryrd}
\usepackage{verbatim}
\usepackage[noadjust]{cite}
\usepackage{graphicx}
\usepackage{tikz}

\theoremstyle{plain}
\newtheorem{thm}{Theorem}
\newtheorem{lem}[thm]{Lemma}

\newtheorem{defn}{Definition}
\newtheorem{example}{Example}
\newtheorem{remark}{Remark}

\usepackage[colorlinks=true]{hyperref}

\begin{document}
\title{Sharper Asymptotically Optimal CDC Schemes via Combinatorial Designs}

\author{Yingjie Cheng, Gaojun Luo, Xiwang Cao, Martianus Frederic Ezerman, and San Ling
\thanks{Y. Cheng, and X. Cao are with the Department of Mathematics, Nanjing University of Aeronautics and Astronautics, Nanjing 210016, China, and also with Key Laboratory of Mathematical Modeling and High Performance Computing of Air Vechicles (NUAA), MIIT, Nanjing 210016, China, e-mails: $\{\rm xwcao,chengyingjie\}$@nuaa.edu.cn}
\thanks{G. Luo, M. F. Ezerman, and S. Ling are with the School of Physical and Mathematical Sciences, Nanyang Technological University, 21 Nanyang Link, Singapore 637371, e-mails: $\{\rm gaojun.luo, fredezerman, lingsan\}$@ntu.edu.sg.}
\thanks{G. Luo, M. F. Ezerman, and S. Ling are supported by Nanyang Technological University Research Grant No. 04INS000047C230GRT01. X. Cao, Y. Cheng, and G. Luo are also supported by the National Natural Science Foundation of China under Grant 12171241.}

}


\maketitle

\begin{abstract}
Coded distributed computing (CDC) was introduced to greatly reduce the communication load for {\tt MapReduce} computing systems. Such a system has $K$ nodes, $N$ input files, and $Q$ {\tt Reduce} functions. Each input file is mapped by $r$ nodes and each {\tt Reduce} function is computed by $s$ nodes. The architecture must allow for coding techniques that achieve the maximum multicast gain. Some CDC schemes that achieve optimal communication load have been proposed before. The parameters $N$ and $Q$ in those schemes, however, grow too fast with respect to $K$ to be of great practical value. To improve the situation, researchers have come up with some asymptotically optimal cascaded CDC schemes with $s+r=K$ from symmetric designs. 

In this paper, we propose new asymptotically optimal cascaded CDC schemes. Akin to known schemes, ours have $r+s=K$ and make use of symmetric designs as construction tools. Unlike previous schemes, ours have much smaller communication loads, given the same set of parameters $K$, $r$, $N$, and $Q$. We also expand the construction tools to include almost difference sets. Using them, we have managed to construct a new asymptotically optimal cascaded CDC scheme.
\end{abstract}

\begin{IEEEkeywords}
Almost difference set, coded distributed computing, communication load, symmetric design.
\end{IEEEkeywords}

\section{Introduction}\label{sec:intro}
Processing large amount of data efficiently is a must in this era of big data. Handling such a computational task lies beyond the capability of a single computer. The challenge to complete huge computational assignments motivates the design of distributed computing systems. The main objective is to greatly expedite task execution by letting distributed computing nodes perform computational jobs in parallel by exploiting the distributed nature of available resources, both computing and storage. It is often the case that a large amount of data needs to be exchanged among the computing nodes, which limits the system's performance. In a Facebook {\tt Hadoop} cluster, for example, it has been observed that $33\%$ of the overall job execution time was spent on data shuffling \cite{chowdhury2011}. We know from \cite{zhang2013} that $70\%$ of the overall job execution time is spent on data shuffling when running a self-join application on an Amazon {\tt EC2} cluster.

S. Li {\it et al.} in \cite{li2017} introduced coded distributed computing(CDC) to reduce the communication load in distributed computing systems. The reduction is the result of CDC's capability to increase the computation load of the so-called Map functions to create novel coding opportunities. Some systems, which had already been in use by then, including Dean and Ghemawat's {\tt MapReduce} \cite{dean2008} and {\tt Spark} of Zaharia {\it et al.} from \cite{zaharia2010}, could subsequently be improved. 

We call a system a $(K,N,r,s,Q)$-CDC when the system has $K$ computing nodes, $N$ input data files of equal size, and $Q$ output values, each of which is computed by a function on the $N$ files. A computation in this system is divided into three phases, namely {\tt Map}, {\tt Shuffle}, and {\tt Reduce}. In the {\tt Map} phase, a given input file is exclusively mapped by a distinct $r$-subset computing nodes to $Q$ \emph{intermediate values} (IVs) with $T$ bits. In the {\tt Shuffle} phase, each {tt Reduce} function is assigned to an $s$-subset of computing nodes. Subsequently, all computing nodes generate coded symbols from their respective local IVs in such a way that each computing node can derive the needed IVs that it cannot, by itself, calculate locally. In the {\tt Reduce} phase, any computing node can compute each {\tt Reduce} function assigned to it after receiving the coded signals during the {\tt Shuffle} phase. We underline the fact that nodes have to spend most of their execution time in exchanging IVs among themselves, causing a substantial communication bottleneck in the system \cite{chowdhury2011}. Hence, it is highly desirable to reduce the execution time in the {\tt Shuffle} phase. 

A fundamental trade-off between \emph{computation load} in the {\tt Map} phase and \emph{communication load} in the {\tt Shuffle} phase was formulated and characterized by Li {\it et al.} in \cite{li2017}. Increasing the computation load by a factor of $r$ can reduce the communication load by the same factor. The authors of the said work also proposed several CDC schemes that achieved the optimal communication load. Their main idea is as follows. In the {\tt Map} phase the nodes need to compute some side information locally. In the {\tt Shuffle} phase, the nodes exchange some coded data among themselves. The side information makes each coded data simultaneously useful for multiple {\tt Reduce} tasks.

In a general $(K,N,r,s,Q)$-CDC scheme, if $s=1$, then each {\tt Reduce} function is calculated by exactly one node. This scheme is similar to the \emph{coded caching scheme} for the D2D network treated in, {\it e.g.}, \cite{li2017} and \cite{ji2015}. If $s\geq 1$, then each {\tt Reduce} function is calculated by multiple nodes. The scheme is known as \emph{cascaded} CDC scheme. Numerous works, {\it e.g.}, \cite{agrawal2020}, \cite{lee2017}, \cite{li2016} and \cite{yan2020}, proposed CDC schemes with \emph{stragglers}. In heterogeneous networks, CDC schemes have been studied in \cite{kiamari2017}, \cite{shakya2018}, \cite{woolsey2021combinatorial}, \cite{woolsey2019}, \cite{woolsey2020coded}, and \cite{xu2019}. Extension to various setups had been pursued. Without attempting a complete listing, we mention works on CDC schemes in wireless network in \cite{li2019} and \cite{li2016edge}, and in the context of matrix multiplication in \cite{lee2017high} and \cite{d2020notes}. Our present work focuses on cascaded CDC schemes. Prior works on such schemes include \cite{cheng2023}, \cite{jiang2022}, \cite{li2017}, \cite{woolsey2021}, and \cite{jiang2020}.

The scheme in \cite{li2017}, henceforth the Li-CDC, splits the data set into $N=\binom{K}{r}$ files and designs $Q=\binom{K}{s}$ output functions, with $r$ being the average number of nodes that store each file and $s$ being the average number of nodes that calculate each function. The Li-CDC achieves the minimum communication load. The number $N=\binom{K}{r}$ of files and the number $Q=\binom{K}{s}$ of functions in the Li-CDC grow too fast with respect to $K$ for practical scenarios. This was shown by Konstantinidis and Ramamoorthy in \cite{kon2020}. Woolsey, Chen, and Ji in \cite{woolsey2021} introduced a combinatorial structure called \emph{hypercube structure} to design the file and function assignments. Their scheme requires the data set to be split into $N=\left(\frac{K}{r}\right)^{r-1}$ files and designs $Q=\left(\frac{K}{r}\right)^{r-1}$ output functions. They also  showed that the communication load of their scheme is close to that of the one in \cite{li2017}. Jiang and Qu in \cite{jiang2020} put forward some cascaded CDC schemes with $N=\left(\frac{K}{r}\right)^{r-1}$ and $Q=\frac{K}{\gcd(K,s)}$ by using placement delivery array. Such an array had previously been introduced by Yan {\it et al.} to construct coded caching schemes in \cite{yan2017}. The communication load of the schemes built by Jiang and Qu, however, is about twice as large as that of the Li-CDC. Recently, Jiang, Wang, and Zhou in \cite{jiang2022} used a symmetric balanced incomplete block design (SBIBD) to generate the data placement and the {\tt Reduce} function assignment to obtain an asymptotically optimal scheme with $K=N=Q$. In \cite{cheng2023}, Cheng, Wu, and Li proposed some asymptotically optimal schemes based on $t$-designs, with $t\geq 2$. More specifically, their main tool consists of $t$-group divisible designs (GDDs). For ease of reference, we list the above-mentioned known cascaded CDC schemes in the first part of Table \ref{table1}. 
\begin{table}
\caption{Previously Known and Our New Cascaded CDC Schemes}
\label{table1}
\renewcommand{\arraystretch}{1.2}
\centering
\begin{tabular}{l|c|c|c|c|c|c|l}
\toprule
Parameters are & Ref. & Nodes & Comput. & Replic. & Input & Output & Communication Load\\
positive integers & &  & Load & Factor & Files & Functions &  \\
\midrule

$1 \leq r, s \leq K$ & \cite{li2017} & $K$ & $r$ & $s$ & $\binom{K}{r}$ & $\binom{K}{s}$ & $\displaystyle{\sum^{\min(r+s,K)}_{\ell=\max(r+1,s)} \frac{\binom{K-r}{K - \ell} \, \binom{r}{\ell - s}} {\binom{K}{s}} \, \frac{\ell-r}{\ell-1}}$ \\

\midrule

$1\leq r\leq K$ & \cite{woolsey2021}& $K$ & $r$ & $r$ & $\left(\frac{K}{r}\right)^{r-1}$ & $\left(\frac{K}{r}\right)^{r-1}$ & $\frac{r^{r} \, (k-r)}{K^{r} \, (r-1)} + $\\
&&&&&&& $\displaystyle{\sum^{r}_{\ell=2} \left(\frac{r}{K}\right)^{r+\ell} \, \binom{r}{\ell} \, \binom{\frac{K}{2}}{2}^{\ell} \, \frac{2^{\ell} \ell}{2 \ell-1}}$ \\
\midrule

$1\leq r,s\leq K$ & \cite{jiang2020} & $K$ & $r$ & $s$ & $\left(\frac{K}{r}\right)^{r-1}$ & $\frac{K}{\gcd(K,s)}$ & $\frac{s}{r-1} \, \left(1-\frac{r}{K}\right)$ \\

\midrule

$(K,r,\lambda)$ SD with & \cite{jiang2022} & $K$ & $r$ & $r$ & $K$ & $K$ & $\frac{r \, (K-r)}{(r-1) \, K}$ \\

\cmidrule{3-8}

$2\leq \lambda \leq r$ & &$K$ &$r$& $K-r$& $K$& $K$& $\frac{K-r}{K-1}$ \\
\midrule

$(N,M,\lambda)$ $t$-design & \cite{cheng2023} & 
$\frac{\lambda \, \binom{N}{t}} {\binom{M}{t}}$ & 
$\frac{\lambda \, \binom{N-1}{t-1}}{\binom{M-1}{t-1}}$ & $\frac{\lambda \, \binom{N-1}{t-1}}{\binom{M-1}{t-1}}$ & $N$ & $N$ & $\frac{N-1}{2N}$ \\

with $t \geq 2$ & &&&&&& \\

\midrule
$(m,q,M,\lambda)$ $t$-GDD & \cite{cheng2023} & 
$\frac{\lambda \, \binom{m}{t} \, q^{t}} {\binom{M}{t}}$ & 
$\frac{\lambda \, \binom{m-1}{t-1} \, q^{t-1}} {\binom{M-1}{t-1}}$ & $\frac{\lambda \, \binom{m-1}{t-1} \, q^{t-1}} {\binom{M-1}{t-1}}$ & $m \, q$ & $m \, q$ & $\frac{1}{2}+\frac{q-2}{m \,q}$\\
\midrule
\midrule
$(K,r,\lambda)$ SD & New & $K$ & $r$ & $K-r$ & $K$ & $K$ & $\frac{(K-1)^{2}-rK+K}{K(K-1)}$ \\
with $2\leq \lambda \leq r$ &&&&&&& \\ 

\midrule
$(n,k,\lambda,\mu)$ ADD & New & $n$ & $k$ & $k$ & $n$ & $n$ & $\frac{n-1}{2n}$ \\
with $1\leq\lambda\leq k-1$ &&&&&&& \\

\midrule
$(n,k,0,\mu)$ ADD & New & $n$ & $k$ & $k$ & $n$ & $n$ & $\frac{2n-2-k(k-1)}{2n}$ \\
\bottomrule
\end{tabular}
\end{table}

This paper has two main contributions. First, we construct a new class of asymptotically optimal schemes for the cascaded case with $r+s=K$. In our construction, we carefully arrange the data placement and assign the {\tt Reduce} functions by using symmetric designs that meet specific requirements. As shown in Table \ref{table1}, our new schemes have the following advantages.
\begin{itemize}
\item Compared with the Li-CDC scheme in \cite{li2017}, our schemes have much smaller $N$ and $Q$. Using the known symmetric designs listed in Table \ref{table0.1}, the respective communication loads of our schemes approximate that of the Li-CDC scheme.
\item Compared with the schemes of Jiang and Qu in \cite{jiang2022}, ours have smaller respective communication loads for the same $(K,r,N,Q)$. Although both our schemes and those in \cite{jiang2022} make use of symmetric designs in their constructions, we devise a different transmission scheme from what Jiang and Qu had chosen. 
\end{itemize}

Second, we present a class of new asymptotically optimal cascaded CDC schemes. They are constructed based on specially built $1$-designs from almost difference sets. Although our schemes bear some similarities with the schemes of Cheng, Wu, and Li in \cite{cheng2023}, the parameters differ, as shown in Table \ref{table1}.

In terms of organization, Section \ref{sec:prelim} introduces useful properties of symmetric designs, almost difference sets, and cascaded CDC systems. We explain two new constructions of CDC schemes in Section \ref{sec:CDC_construct}. Comparative performance analysis of our new schemes relative to known schemes can be found in Section \ref{sec:perform}. The section also collects concluding remarks.

\section{Preliminaries}\label{sec:prelim}
We denote by $|\cdot|$ the cardinality of a set or the length of a vector. For any positive integers $a$ and $b$ with $a<b$, we use  $[a,b]$ to denote the set $\{a,a+1,\ldots,b\}$. If $a=1$, then we use the shorter form $[b]$. 

\subsection{Cascaded Coded Distributed Computing Systems}
\begin{figure}[b!]
\centering
\caption{A two-stage distributed computing framework. The overall computation is decomposed into computing two sets of functions, namely map and reduce functions, consecutively.}
\label{fig1}
\includegraphics[width=0.8\linewidth]{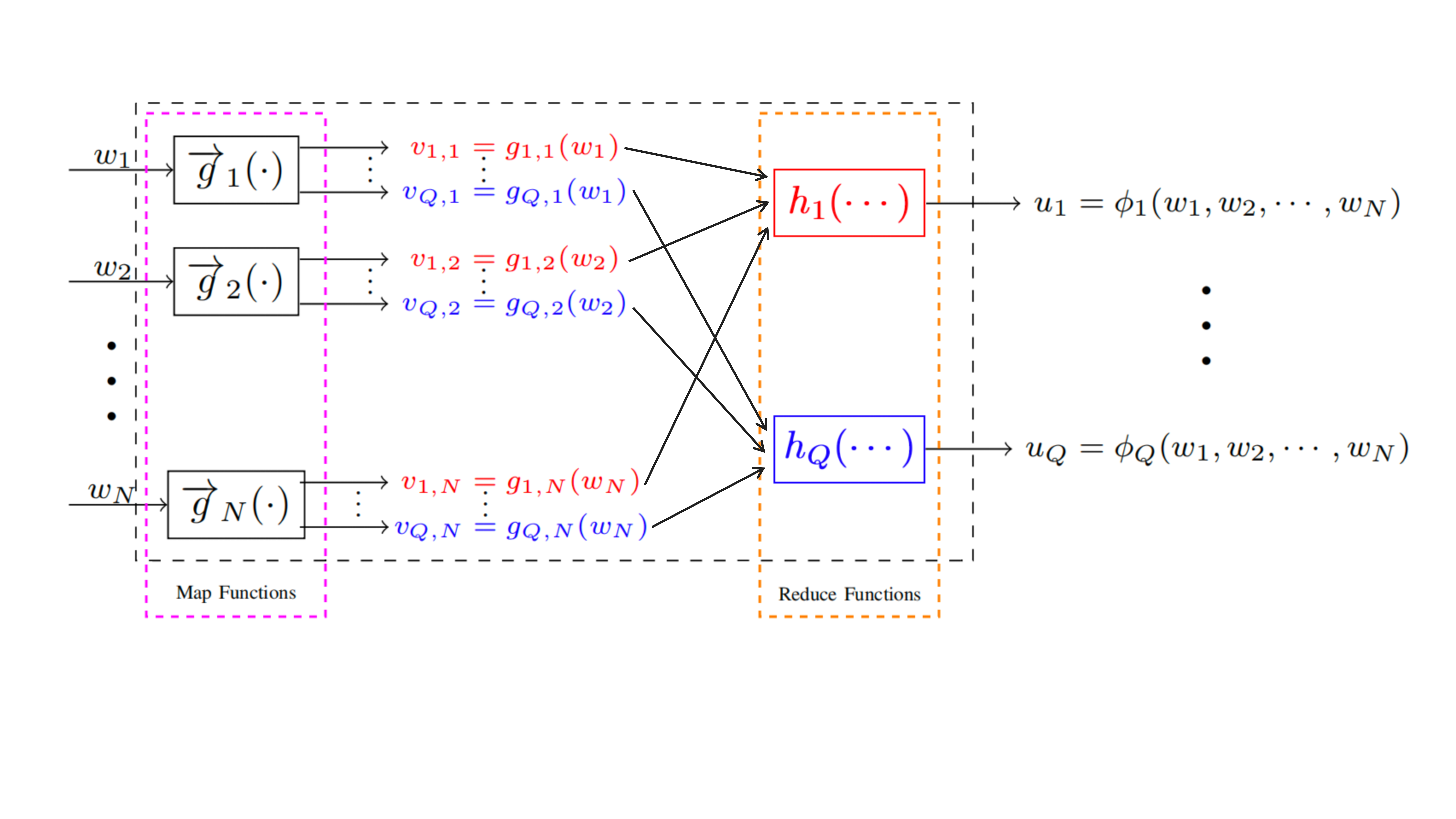}
\end{figure}

In a coded distributed computing system, $K$ distributed computing nodes compute $Q$ {\tt Reduce} functions by taking advantage of $N$ input files, each of equal size. Let $W=\{w_{1},w_{2},\ldots,w_{N}\}$ be the set of the $N$ files, each of size $B$ bits. The set of functions is $\mathcal{Q}=\{\phi_{1},\phi_{2},\ldots,\phi_{Q}\}$, where, for any $q\in[Q]$, $\phi_{q}$ maps the $N$ files to a $C$-bit value $u_{q}:=\phi_{q}(w_{1},w_{2},\ldots,w_{N}) \in \mathbb{F}_{2^{C}}$. Figure \ref{fig1} depicts how each output function $\phi_{q}$ is decomposed into 
\[
\phi_{q}(w_{1},w_{2},\ldots,w_{N}) = h_{q}(g_{q,1}(w_{1}),g_{q,2}(w_{2}),\ldots,g_{q,N}(w_{N})).
\]
Here, $g_{q,n}$ is a {\tt Map} function for any $q\in[Q]$ and $n\in[N]$, whereas $h_{q}$ is a {\tt Reduce} function for any $q\in[Q]$. We name $v_{q,n} := g_{q,n}(w_{n}) \in \mathbb{F}_{2^{T}}$, where $q\in[Q]$ and $n\in[N]$, an \emph{intermediate value} (IV) of length $T$. Figure \ref{fig1} shows that a cascaded CDC consists of three phases.

\begin{itemize}
\item {\tt Map} Phase: Each node $k\in \mathcal{K}$ stores $M$ files. For each file $w_{n}$, let $\mathcal{D}_{n}$ represent the set of nodes, each of which stores file $w_{n}$. We can then write the files stored by node $k$ as elements in the set
\begin{equation} \label{eq0.1}
\mathcal{Z}_{k}=\{w_{n} : n\in[N],k\in \mathcal{D}_{n}\}.
\end{equation}
Using the stored files in (\ref{eq0.1}) and the {\tt Map} functions in $\{g_{q,n}(\cdot) : q \in [Q] \mbox { and } n \in [N]\}$, node $k$ can compute the IVs in
\[
\mathcal{I}_{k}=\{v_{q,n}=g_{q,n}(w_{n}) \in \mathbb{F}_{2^{T}}: q\in[Q],n\in[N],k\in \mathcal{D}_{n}\}.
\]

\item {\tt Shuffle} Phase: For any $k\in \mathcal{K}$, let
\begin{equation} \label{eq0.2}
\mathcal{Q}_{k}=\{\phi_{q} : q\in[Q],k\in \mathcal{A}_{q}\}
\end{equation}
be the set of output functions to be calculated by node $k$. Collectively and in a coordinated way, the nodes exchange calculated IVs such that each node can derive the IVs that it cannot locally calculated. The node $k$, for $k\in \mathcal{K}$, multicasts a coded message $X_{k}$ of length $\ell_{k}$.

\item {\tt Reduce} Phase: Upon receiving the coded signals $\mathcal{X}=\{X_{1},X_{2},\ldots,X_{K}\}$ and its locally computed IVs in $\mathcal{I}_{k}$, node $k\in \mathcal{K}$ can compute each {\tt Reduce} function in $\mathcal{Q}_{k}$.
\end{itemize}

Keeping the relevant definitions from \cite{li2017}, we know that there are two important quantities that measure the goodness of a CDC. First, the average number of nodes that store each file is the \emph{computation load} 
\[
r=\frac{\sum^{K}_{k=1} |\mathcal{W}_{k}|}{N}.
\]
Second, the ratio of the amount of transmitted data to the product $Q \, N \, T$ is the \emph{communication load}
\[
L=\frac{\sum^{K}_{k=1} \ell_{k}}{Q \, N \, T}.
\]

\begin{lem} \label{lem0.1} \cite{li2017}
Let $K \in \mathbb{N}$. Given $r,s\in[K]$, there exists a CDC scheme that achieves the optimal communication load
\begin{equation}\label{eq:optcomm}
L=\sum^{\min(r+s,K)}_{\ell = \max(r+1,s)} 
\frac{\binom{K-r}{K-\ell} \, \binom{r}{\ell-s}}{\binom{K}{s}} \, \frac{\ell-r}{\ell-1},
\end{equation}
where $r$ is the computation load and $s$ is the number of nodes that calculate each function.
\end{lem}

Given $K$, $r$, and $s$, we call any CDC scheme whose $L$ is as in \eqref{eq:optcomm} a 
\emph{Li-CDC scheme} and denote by $L_{\rm Li}$ the communication load of a Li-CDC scheme. It is clear from Lemma \ref{lem0.1} that, given $r$ and $s$, we seek to minimize $L$.

\subsection{Almost Difference Sets}
We recall useful results on almost difference sets. Let $(A,+)$ be a finite abelian group of order $n$ and let $D$ be a subset of size $k$ in $(A,+)$. The {\tt difference} function on a subset $D$ of $(A,+)$ is
\begin{equation} \label{eq1.1}
{\rm diff}_{D}(x) = |D\cap(D+x)|,
\end{equation}
where $D+x=\{y+x:y\in D\}$ and $x\in A$. 
\begin{defn} \cite{ding2014}
Let $(A,+)$ be an abelian group of order $n$. A $k$-subset $D$ of $A$ is an $(n,k,\lambda,t)$ almost difference set (ADS) of $A$ if ${\rm diff}_{D}(x)$ takes on $\lambda$ altogether $t$ times and $\lambda+1$ altogether $n-1-t$ times as $x$ traverses the nonzero elements of $A$.
\end{defn}
\begin{remark} 
We list useful facts from \cite{ding2014}. Let an abelian group $(A,+)$ be given.
\begin{enumerate}
\item If an $(n,k,\lambda,t)$ ADS exists, then
$k(k-1)=t\lambda+(n-1-t)(\lambda+1)$.
\item If $D$ is an $(n,k,\lambda,t)$ ADS, then its complement $D^{c}=A \setminus D$ is an $(n,n-k,n-2k+\lambda,t)$ ADS in $(A,+)$.
\item If $D$ is an $(n,k,0,t)$ ADS with 
$t=n-1-k(k-1)$, then $D$ is also called a \emph{modular Golomb ruler} in $(A,+)$.
\end{enumerate}
\end{remark}

Ruzsa introduced a class of modular Golomb ruler, which we will use in the next section, in \cite{ruzsa1993}.

\begin{lem} \cite{ruzsa1993} \label{lem0.2}
For every prime $p$, there exists a $(p^{2}-p,p-1,0,2p-3)$ almost difference set. The missing differences are the $2p-2$ multiples of $p$ or $p-1$.
\end{lem}

\subsection{Symmetric Designs}
We gather useful results on relevant combinatorial designs. 
\begin{defn} \cite{ionin2006}
Let $\mathcal{X}$ be a set of $v$ elements. Let $\mathcal{B}:=\{B_{1},B_{2},\ldots,B_{u}\}$ be such that $B_{i}\subseteq \mathcal{X}$ and $|B_{i}| =t$ for any $i\in[u]$. Given any two distinct elements $x,y\in\mathcal{X}$, if there exist exactly $\lambda$ elements in $\mathcal{B}$ containing them, then $(\mathcal{X},\mathcal{B})$ is a $(v,t,\lambda)$ balanced incomplete block design (BIBD). For each $i\in[u]$, we call $B_{i}$ a \emph{block}.
\end{defn}

\begin{remark} 
Since any BIBD is also a $2$-design, we have 
$|\mathcal{B}|=\displaystyle{\frac{\lambda \, v \, (v-1)} {t \,(t-1)}}$ for any $(v,t,\lambda)$ BIBD $(\mathcal{X},\mathcal{B})$.
\end{remark}

We will soon use symmetric designs as a main construction tool for a class of CDC schemes.

\begin{defn} \cite{ionin2006}
A $(v,t,\lambda)$ BIBD $(\mathcal{X},\mathcal{B})$ is a $(v,t,\lambda)$ symmetric design (SD) if $ |\mathcal{B}|=v$.
\end{defn}

\begin{lem} \label{lem1.1} \cite{ionin2006}
Given a $(v,t,\lambda)$ symmetric design $(\mathcal{X},\mathcal{B})$, the following statements hold.
\begin{enumerate}
	\item Each $x\in \mathcal{X}$ is contained in $t$ blocks among the elements of $\mathcal{B}$.
	\item For any two distinct blocks $B$ and $B'$ in $\mathcal{B}$, we have $| B\bigcap B'|=\lambda$.
	\item $\lambda=\frac{t \,(t-1)}{v-1}$.
\end{enumerate}
\end{lem}
In \cite{ionin2006}, Ionin and van Trung listed the parameters of four known classes of symmetric designs.

\begin{lem} \cite{ionin2006}
A $(v,k,\lambda)$ symmetric design exists if its parameters can be found in Table \ref{table0.1}.
\end{lem}
\begin{table}[h!]
\caption{Known parameters of $(v,k,\lambda)$ symmetric designs}
\label{table0.1}
\renewcommand{\arraystretch}{1.1}
\centering
\begin{tabular}{l|l|l|l}
\toprule
Conditions & $v$ & $t$ & $\lambda$  \\
\midrule
$b$ is a prime power & $b^{2}+b+1$ & $b+1$ & $1$   \\
$b$ is a prime power & $b^{3}+b^{2}+b+1$ & $b^{2}+b+1$ & $b+1$   \\
$b$ is a prime power & $b^{3}+2b^{2}$ & $b^{2}+b$ & $b$   \\
$b-1$ and $b^{2}-b+1$ are prime powers & $b^{3}+b+1$ & $b^{2}+1$ & $b$   \\
\bottomrule
\end{tabular}
\end{table}
The four classes of symmetric designs in the table play important roles in our construction of asymptotically optimal cascaded CDC schemes. 

\section{Two Constructions of CDC Schemes}\label{sec:CDC_construct}

\subsection{Construction One}\label{subsec:scheme1}
This subsection introduces a new construction method for the case $r\neq s$. Let $(\mathcal{X},\mathfrak{B})$ be an $(N,t,\lambda)$ symmetric design, $\mathcal{X}=\{x_{1},x_{2},\ldots,x_{N}\}$, and $\mathfrak{B} = \{\mathcal{B}_{1},\mathcal{B}_{2},\ldots,\mathcal{B}_{N}\}$. By Lemma \ref{lem1.1}, any two distinct blocks $\mathcal{B}_{i}$ and $\mathcal{B}_{j}$ intersect in exactly $\lambda$ points. We now construct a CDC scheme with $N$ nodes, where $\mathcal{K} = \mathcal{B}$, on $N$ files, which are elements of $\mathcal{W} = \{w_{x_{1}},w_{x_{2}},\ldots,w_{x_{N}}\}$, and $N$ functions in $\mathcal{Q} = \{\phi_{x_{1}},\phi_{x_{2}},\ldots,\phi_{x_{N}}\}$. Each node stores $t$ files and each {\tt Reduce} function is computed by $s$ nodes.

During the {\tt Map} phase, each node $\mathcal{B}\in \mathfrak{B}$ stores the files in the set $Z_{\mathcal{B}}=\{w_{x} : x\in\mathcal{B}, \ x\in\mathcal{X}\}$. Since $|\mathcal{B}|=t$, for any block $B$, the computation load is
\[
r=\frac{\sum^{N}_{i=1} |Z_{i}|}{N}=\frac{t \, N}{N}=t.
\]
In the {\tt Shuffle} phase, we arrange each node $\mathcal{B}\in \mathfrak{B}$ to compute the {\tt Reduce} functions
\[
\mathcal{Q}_{\mathcal{B}}= \left\{u_{y} = \phi_{y}(w_{x_{1}},w_{x_{2}},\ldots,w_{x_{N}}\right) : y\in \mathcal{X}, \, y\in \mathcal{\overline{B}}\},
\]
with $\mathcal{\overline{B}}$ denoting the complement set of $\mathcal{B}$ with respect to $\mathcal{X}$. Given the assigned stored files and the set $\mathcal{Q}_{\mathcal{B}}$, node $\mathcal{B}$ can compute the intermediate values in the set
\[
\mathcal{I}_{\mathcal{B}} = \{v_{y,x} = g_{y,x}(w_{x}) : x,y\in\mathcal{X}, \ x\in\mathcal{B}\}.
\]
Hence, for any $x,y\in\mathcal{X}$ and for any block $\mathcal{B}\in \mathfrak{B}$, the intermediate value $v_{y,x}$ is both required and cannot be locally computed by node $\mathcal{B}$ if and only if $y \in \mathcal{\overline{B}}$ and $x \in \mathcal{\overline{B}}$, {\it i.e.}, $y \notin \mathcal{B}$ and $x \notin \mathcal{B}$. The intermediate value $v_{y,x}$ is \emph{locally computable} by node $\mathcal{B}$ if and only if $x\in \mathcal{B}$. 

Based on what we have just investigated, we can divide the delivery strategy into two classes. The first class is for the $N$ intermediate values $v_{x,x} : x \in \mathcal{X}$. We cluster each $v_{x,x}$ into $t$ segments as
\[
v_{x,x} = \left(v^{\mathcal{B}_{k_{1}}}_{x,x}, v^{\mathcal{B}_{k_{2}}}_{x,x}, \ldots, v^{\mathcal{B}_{k_{t}}}_{x,x}\right),
\]
where $x \in \mathcal{B}_{k_{i}}$ for each $i\in [t]$. Since $v_{x,x} \in \mathbb{F}_{2^{T}}$, we know that $v^{\mathcal{B}_{k_{i}}}_{x,x} \in \mathbb{F}_{2^{\frac{T}{t}}}$.

A node $\mathcal{B}_{k}$ has access to $t$ stored files in  $\{w_{z_{1}},w_{z_{2}},\ldots,w_{z_{t}}\}$, giving it the intermediate values in 
\[
\mathcal{V}_{k} = \left\{v_{w_{z_{1}},w_{z_{1}}},v_{w_{z_{2}},w_{z_{2}}}, \ldots,v_{w_{z_{t}},w_{z_{t}}}\right\}.
\]
If $\alpha_{1},\alpha_{2},\ldots,\alpha_{t} \in \mathbb{F}_{2^{\frac{T}{t}}}$ are all distinct, then $t$ must be a divisor of $T$ and $T \geq t^{2}$. Node $\mathcal{B}_{k}$ multicasts the $t-\lambda$ signals
\begin{align*}
X^{\mathcal{B}_{k}}[1] &= v^{\mathcal{B}_{k}}_{w_{z_{1}},w_{z_{1}}}+ v^{\mathcal{B}_{k}}_{w_{z_{2}},w_{z_{2}}}+\ldots+ v^{\mathcal{B}_{k}}_{w_{z_{t}},w_{z_{t}}},\\
X^{\mathcal{B}_{k}}[2] &=\alpha_{1}v^{\mathcal{B}_{k}}_{w_{z_{1}},w_{z_{1}}}+ \alpha_{2}v^{\mathcal{B}_{k}}_{w_{z_{2}},w_{z_{2}}}+ \ldots+\alpha_{t}v^{\mathcal{B}_{k}}_{w_{z_{t}},w_{z_{t}}},\\
&\vdots \\
X^{\mathcal{B}_{k}}[t-\lambda] 
& =\alpha^{t-\lambda-1}_{1}v^{\mathcal{B}_{k}}_{w_{z_{1}}, w_{z_{1}}}+\alpha^{t-\lambda-1}_{2}v^{\mathcal{B}_{k}}_{w_{z_{2}}, w_{z_{2}}}+\ldots+ \alpha^{t-\lambda-1}_{t}v^{\mathcal{B}_{k}}_{w_{z_{t}},w_{z_{t}}},
\end{align*}
which we express as 
\[
\begin{pmatrix}
X^{\mathcal{B}_{k}}[1] \\
X^{\mathcal{B}_{k}}[2] \\
\vdots \\
X^{\mathcal{B}_{k}}[t-\lambda]
\end{pmatrix}
= 
\begin{pmatrix}
1 & 1 & \cdots & 1 \\
\alpha_{1} & \alpha_{2} & \cdots & \alpha_{t} \\
\vdots & \vdots & \ddots & \vdots\\
\alpha^{t-\lambda-1}_{1} & \alpha^{t-\lambda-1}_{2} & \cdots & \alpha^{t-\lambda-1}_{t} \\
\end{pmatrix} ~
\begin{pmatrix}
v^{\mathcal{B}_{k}}_{w_{z_{1}},w_{z_{1}}} \\
v^{\mathcal{B}_{k}}_{w_{z_{2}},w_{z_{2}}} \\
\vdots \\
v^{\mathcal{B}_{k}}_{w_{z_{t}},w_{z_{t}}}
\end{pmatrix}.
\]
The total number of bits transmitted by $\mathcal{B}_{k}$ is, therefore, $(t-\lambda)\frac{T}{t}$, which comes from $(t-\lambda)\frac{1}{t}$ intermediate values. Thus, the total number of intermediate values transmitted by all the nodes combined is $(t-\lambda)\frac{v}{t}$.

If a node $\mathcal{B}$ is unable to compute $v_{y,y}$, then $w_{y} \notin \mathcal{B}$. Hence, there exist nodes $\mathcal{B}_{u_{i}} : i\in[t]$ such that $w_{y} \in \mathcal{B}_{u_{i}}$. Without loss of generality, let $\mathcal{B}_{u_{1}}$ be a node whose stored files are in $\left\{w_{y},w_{\ell_{1}},w_{\ell_{2}},\ldots,w_{\ell_{t-1}}\right\}$. By Lemma \ref{lem1.1}, we have $|\mathcal{B}_{u_{1}}\bigcap \mathcal{B}|=\lambda$. If these $\lambda$ stored files are elements of $\left\{w_{\ell_{t-\lambda}},w_{\ell_{t-\lambda+1}},\ldots, w_{\ell_{t-1}}\right\}$, then node $\mathcal{B}$ can locally compute 
\[
v^{\mathcal{B}_{u_{1}}}_{w_{\ell_{t-\lambda}}, w_{\ell_{t-\lambda}}},   \quad v^{\mathcal{B}_{u_{1}}}_{w_{\ell_{t-\lambda+1}},w_{\ell_{t-\lambda+1}}}, \quad \ldots, \quad v^{\mathcal{B}_{u_{1}}}_{w_{\ell_{t-1}},w_{\ell_{t-1}}}.
\]
Thus, $\mathcal{B}$ only needs to solve the system of equations
\[
\begin{pmatrix}
X^{\mathcal{B}_{u_{1}}}[1]- \sum^{t}_{i=t-\lambda+1} v^{\mathcal{B}_{u_{1}}}_{w_{l_{i-1}},w_{l_{i-1}}} \\
X^{\mathcal{B}_{u_{1}}}[2]-\sum^{t}_{i=t-\lambda+1}\alpha_{i}v^{\mathcal{B}_{u_{1}}}_{w_{l_{i-1}},w_{l_{i-1}}} \\
\vdots \\
X^{\mathcal{B}_{u_{1}}}[t-\lambda]-\sum^{t}_{i=t-\lambda+1}\alpha^{t-\lambda-1}_{i}v^{\mathcal{B}_{u_{1}}}_{w_{l_{i-1}},w_{l_{i-1}}}
\end{pmatrix}
=
\begin{pmatrix}
1 & 1 & \cdots & 1 \\
\alpha_{1} & \alpha_{2} & \cdots & \alpha_{t-\lambda} \\
\vdots & \vdots & \ddots & \vdots\\
\alpha^{t-\lambda-1}_{1} & \alpha^{t-\lambda-1}_{2} & \cdots & \alpha^{t-\lambda-1}_{t-\lambda} \\
\end{pmatrix} ~
\begin{pmatrix}
v^{\mathcal{B}_{u_{1}}}_{w_{y},w_{y}} \\
v^{\mathcal{B}_{u_{1}}}_{w_{l_{1}},w_{l_{1}}} \\
\vdots \\
v^{\mathcal{B}_{u_{1}}}_{w_{l_{t-\lambda-1}},w_{l_{t-\lambda-1}}}
\end{pmatrix}.
\]
The coefficient matrix is clearly Vandermonde. Since $\alpha_{1},\alpha_{2},\ldots,\alpha_{t}$ are all distinct, node $\mathcal{B}$ decodes $v^{\mathcal{B}_{u_{1}}}_{w_{y},w_{y}}$ for node $\mathcal{B}_{u_{1}}$. Similarly, node $\mathcal{B}$ can also derive $v^{\mathcal{B}_{u_{i}}}_{w_{y},w_{y}}$ for any node $\mathcal{B}_{u_{i}} : i\in\{2,3,\ldots,t\}$. Thus, node  $\mathcal{B}$ can derive
\[
v_{w_{y},w_{y}} = \left\{v^{\mathcal{B}_{u_{1}}}_{w_{y}, w_{y}},v^{\mathcal{B}_{u_{2}}}_{w_{y},w_{y}}, \ldots, v^{\mathcal{B}_{u_{t}}}_{w_{y},w_{y}}\right\}.
\]

Proceeding to the second class of intermediate values $v_{x,y}$, where $x, y \in\mathcal{X}$ are distinct, we cluster $v_{x,y}$ into the $\lambda$ segments
\[
v_{x,y}= \left(v^{\mathcal{B}_{s_{1}}}_{x,y}, v^{\mathcal{B}_{s_{2}}}_{x,y},\ldots, v^{\mathcal{B}_{s_{\lambda}}}_{x,y}\right),
\]
where $x,y\in \mathcal{B}_{s_{i}}$ for any $i\in [\lambda]$. Since $v_{x,y}\in\mathbb{F}_{2^{T}}$, it is immediate to confirm that $v^{\mathcal{B}_{s_{i}}}_{x,y}\in \mathbb{F}_{2^{\frac{T}{\lambda}}}$ for any $i\in [\lambda]$.

Any node $\mathcal{B}_{s}$ has access to $t$ stored files in $\{w_{a_{1}},w_{a_{2}},\ldots,w_{a_{t}}\}$. Hence, $\mathcal{B}_{s}$ has the intermediate values in 
\[
\left\{v_{w_{a_{1}},w_{a_{2}}}, v_{w_{a_{1}},w_{a_{3}}}, \ldots,v_{w_{a_{1}},w_{a_{t}}},\ldots,v_{w_{a_{t}}, w_{a_{1}}},v_{w_{a_{t}},w_{a_{2}}},\ldots, v_{w_{a_{t}},w_{a_{t-1}}}\right\}.
\]
If $\beta_{1},\beta_{2},\ldots,\beta_{t-1}\in \mathbb{F}_{2^{\frac{T}{\lambda}}}$ are all distinct, then $\lambda$ divides $T$ and $T \geq \lambda(t-1)$. The $t \, (t-\lambda-1)$ signals that node $\mathcal{B}_{k}$ multicasts can be expressed as
\[
\begin{pmatrix}
Y_{i}^{\mathcal{B}_{s}}[1] \\
Y_{i}^{\mathcal{B}_{s}}[2] \\
\vdots \\
Y_{i}^{\mathcal{B}_{s}}[t-\lambda-1]
\end{pmatrix}
= 
\begin{pmatrix}
1 & 1 & \cdots & 1 \\
\beta_{1} & \beta_{2} & \cdots & \beta_{t-1} \\
\vdots & \vdots & \ddots & \vdots\\
\beta^{t-\lambda-2}_{1} & \beta^{t-\lambda-2}_{2} & \cdots & \beta^{t-\lambda-2}_{t-1} \\
\end{pmatrix} 
\begin{pmatrix}
v^{\mathcal{B}_{s}}_{w_{a_{i}},w_{a_{1}}} \\
v^{\mathcal{B}_{s}}_{w_{a_{i}},w_{a_{2}}} \\
\vdots \\
v^{\mathcal{B}_{s}}_{w_{a_{i}},w_{a_{t}}}
\end{pmatrix},
\]
where $i\in [t]$. The total number of bits transmitted by $\mathcal{B}_{s}$ is, therefore, $t \, (t-\lambda-1) \, \frac{T}{\lambda}$, which comes from $t \, (t-\lambda-1) \, \frac{1}{\lambda}$ intermediate values. Thus, the total number of intermediate values $v_{x,y}$ transmitted by all nodes combined is $t \, (t-\lambda-1) \, \frac{v}{\lambda}$.

If a node $\mathcal{B}_{m}$ is unable to compute $v_{x,y}$, then $w_{x},w_{y} \notin\mathcal{B}_{m}$. Since $(\mathcal{X},\mathfrak{B})$ is a symmetric design, there exist $\lambda$ nodes $\mathcal{B}_{n_{i}} : i\in[\lambda]$ with access to files $w_{x}$ and $w_{y}$. Without loss of generality, let $\mathcal{B}_{n_{1}}$ be a node such that its stored files are the elements in $\{w_{x},w_{y},w_{b_{1}},w_{b_{2}},\ldots,w_{b_{t-2}}\}$. By Lemma \ref{lem1.1}, $|\mathcal{B}_{n_{1}}\bigcap\mathcal{B}_{m}|=\lambda$. If the $\lambda$ stored files are the elements in $\{w_{b_{t-\lambda-1}}, w_{b_{t-\lambda}}, \ldots,w_{b_{t-2}}\}$, then node $\mathcal{B}$ can locally compute 
\[
v^{\mathcal{B}_{n_{1}}}_{w_{x}, w_{b_{t-\lambda-1}}}, \quad v^{\mathcal{B}_{n_{1}}}_{w_{x}, w_{b_{t-\lambda}}}, \quad \ldots, \quad v^{\mathcal{B}_{n_{1}}}_{w_{x},w_{b_{t-2}}}.
\]
Thus, $\mathcal{B}_{m}$ only needs to solve the system of equations
\begin{align*}
\begin{pmatrix}
Y_{x}^{\mathcal{B}_{n_{1}}}[1] - \sum^{t}_{i=t-\lambda+1} v^{\mathcal{B}_{n_{1}}}_{w_{x},w_{b_{i-2}}} \\
Y_{x}^{\mathcal{B}_{n_{1}}}[2] - \sum^{t}_{i=t-\lambda+1}\beta_{i-1} v^{\mathcal{B}_{n_{1}}}_{w_{x},w_{b_{i-2}}} \\
\vdots \\
Y_{x}^{\mathcal{B}_{n_{1}}}[t-\lambda-1]- \sum^{t}_{i=t-\lambda+1}\beta^{t-\lambda-2}_{i-1} v^{\mathcal{B}_{n_{1}}}_{w_{x},w_{b_{i-2}}}
\end{pmatrix}
 = 
\begin{pmatrix}
1 & 1 & \cdots & 1 \\
\beta_{1} & \beta_{2} & \cdots & \beta_{t-\lambda-1} \\
\vdots & \vdots & \ddots & \vdots\\
\beta^{t-\lambda-2}_{1} & \beta^{t-\lambda-2}_{2} & \cdots & \beta^{t-\lambda-2}_{t-\lambda-1}
\end{pmatrix}
\begin{pmatrix}
v^{\mathcal{B}_{n_{1}}}_{w_{x},w_{y}} \\
v^{\mathcal{B}_{n_{1}}}_{w_{x},w_{b_{1}}} \\
\vdots \\
v^{\mathcal{B}_{n_{1}}}_{w_{x},w_{b_{t-2}}}
\end{pmatrix}.
\end{align*}
The coefficient matrix is obviously Vandermonde. Since $\beta_{1},\beta_{2},\ldots,\beta_{t-1}$ are all distinct, node $\mathcal{B}_{m}$ decodes $v^{\mathcal{B}_{n_{1}}}_{w_{x},w_{y}}$ for node $\mathcal{B}_{n_{1}}$. Similarly, node $\mathcal{B}_{m}$ can provide $v^{\mathcal{B}_{n_{i}}}_{w_{x},w_{y}}$ to any node $\mathcal{B}_{n_{i}} : i\in\{2,3,\ldots,\lambda\}$. Thus, node $\mathcal{B}_{m}$ can derive
\[
v_{w_{x},w_{y}}= \left(v^{\mathcal{B}_{n_{1}}}_{w_{x},w_{y}}, v^{\mathcal{B}_{n_{2}}}_{w_{x},w_{y}},\ldots,
v^{\mathcal{B}_{n_{t}}}_{w_{x},w_{y}}\right).
\]

Since $\lambda=\frac{t(t-1)}{v-1}$ in the known $(v,t,\lambda)$ SD,  the communication load is
\begin{align*}
L &= \frac{t(t-1-\lambda)\frac{Tv}{\lambda} + \frac{vT}{t}(t-\lambda)}{Q \, N \, T} 
= \frac{t(t-1-\lambda) \frac{Tv}{\lambda} + \frac{vT}{t}(t-\lambda)}{v^{2}T} \\
& = \frac{t(t-1-\frac{t(t-1)}{v-1}) \frac{v-1}{t(t-1)} + \frac{1}{t}(t-\frac{t(t-1)}{v-1})}{v} = \frac{(v-1)^{2}-t(v-1)+v-1-t+1}{v(v-1)} \\
&= \frac{(v-1)^{2}-tv+v}{v(v-1)}.
\end{align*}

In the {\tt Reduce} phase, we know that each node $\mathcal{B}\in \mathfrak{B}$ can derive the intermediate values 
\[
\{v_{x,y} : x,y\in\mathcal{X}, \ x,y\in \mathcal{\overline{B}}\}
\]
during the {\tt Shuffle} phase. Node $\mathcal{B}$ can locally compute the {\tt Reduce} functions
\[
\mathcal{Q}_{\mathcal{B}} = \left\{u_{y} = \phi_{y}(w_{x_{1}},w_{x_{2}},\ldots,w_{x_{N}}) : y\in \mathcal{X}, \ y\in \mathcal{\overline{B}}\right\}.
\]
We formalize the above discussions in the following theorem.
\begin{thm} \label{thm2.3}
Given a $(v,t,\lambda)$ SD with $t>\lambda+1$, one can construct a CDC scheme with $v$ distributed computing nodes, $N=v$ files and $Q=v$ output functions such that 
\begin{enumerate}
	\item each output function is computed by $s=v-t$ nodes,
	\item the computation load is $r=t$, and
	\item the communication load is $L=\frac{(v-1)^{2}-tv+v}{v(v-1)}$.
\end{enumerate}
\end{thm}

We use $(7,3,1)$ SD in an example to illustrate our construction.
\begin{example}
\begin{table}[b!]
	\caption{Intermediate values $\{v_{q,n}\}$ required by nodes in $\mathfrak{B}$}
	\label{table2.1}
	\renewcommand{\arraystretch}{1.1}
	\centering
	\begin{tabular}{c|ccccccc}
		\toprule
		Parameters & \multicolumn{7}{c}{Node Set $\mathfrak{B}$} \\
		\midrule
		 & $\mathcal{B}_{1}$ & $\mathcal{B}_{2}$ & $\mathcal{B}_{3}$ & $\mathcal{B}_{4}$ & $\mathcal{B}_{5}$ & $\mathcal{B}_{6}$ & $\mathcal{B}_{7}$  \\
		\midrule
		$q$ & $3,5,6,7$ & $1,4,6,7$ & $1,2,5,7$ & $1,2,3,6$ & $2,3,4,7$ & $1,3,4,5$ & $2,4,5,6$  \\
		\midrule
		$n$ & $3,5,6,7$ & $1,4,6,7$ & $1,2,5,7$ & $1,2,3,6$ & $2,3,4,7$ & $1,3,4,5$ & $2,4,5,6$  \\
		\bottomrule
	\end{tabular}
\end{table}

When $N=Q=K=7$, there are $7$ files in $\mathcal{W} = \{w_{1},w_{2},\ldots,w_{7}\}$ and $7$ functions in $\mathcal{Q}= \{\phi_{1}, \phi_{2},\ldots,\phi_{7}\}$. In the first stage, the nodes store the respective files
\begin{align*}
\mathcal{Z}_{\mathcal{B}_{1}} &=\{w_{1},w_{2},w_{4}\}, 
& \mathcal{Z}_{\mathcal{B}_{2}} &=\{w_{2},w_{3},w_{5}\}, 
& \mathcal{Z}_{\mathcal{B}_{3}} &=\{w_{3},w_{4},w_{6}\}, 
& \mathcal{Z}_{\mathcal{B}_{4}} &=\{w_{4},w_{5},w_{7}\},\\
\mathcal{Z}_{\mathcal{B}_{5}}&=\{w_{1},w_{5},w_{6}\}, 
& \mathcal{Z}_{\mathcal{B}_{6}} &=\{w_{2},w_{6},w_{7}\}, 
& \mathcal{Z}_{\mathcal{B}_{7}} &=\{w_{1},w_{3},w_{7}\}. 
\end{align*}
The computation load is $r=\frac{3 \cdot 7}{7}=3$.

If the {\tt Reduce} functions are arranged by nodes as 
\begin{align*}
\mathcal{Q}_{\mathcal{B}_{1}} &=\{\phi_{3},\phi_{5},\phi_{6},\phi_{7}\}, 
&\mathcal{Q}_{\mathcal{B}_{2}} &=\{\phi_{1},\phi_{4},\phi_{6},\phi_{7}\}, &\mathcal{Q}_{\mathcal{B}_{3}} &=\{\phi_{1},\phi_{2},\phi_{5},\phi_{7}\}, 
&\mathcal{Q}_{\mathcal{B}_{4}} &=\{\phi_{1},\phi_{2},\phi_{3},\phi_{6}\},\\
\mathcal{Q}_{\mathcal{B}_{5}}&=\{\phi_{2},\phi_{3},\phi_{4},\phi_{7}\}, &\mathcal{Q}_{\mathcal{B}_{6}} &=\{\phi_{1},\phi_{3},\phi_{4},\phi_{5}\}, &\mathcal{Q}_{\mathcal{B}_{7}}&=\{\phi_{2},\phi_{4},\phi_{5},\phi_{6}\},
\end{align*}
then each function is computed by $s=4$ nodes.

The locally computable intermediate values, arranged by nodes, can be listed as
\begin{align*}
\mathcal{I}_{\mathcal{B}_{1}} &=\{v_{q,n} : q\in[7], n\in\{1,2,4\}\}, 
&\mathcal{I}_{\mathcal{B}_{2}} &=\{v_{q,n} : q\in[7], n\in\{2,3,5\}\}, 
&\mathcal{I}_{\mathcal{B}_{3}} &=\{v_{q,n} : q\in[7], n\in\{3,4,6\}\}, \\
\mathcal{I}_{\mathcal{B}_{4}} &=\{v_{q,n} : q\in[7], n\in\{4,5,7\}\},
&\mathcal{I}_{\mathcal{B}_{5}}&=\{v_{q,n} : q\in[7], n\in\{1,5,6\}\}, &\mathcal{I}_{\mathcal{B}_{6}}&=\{v_{q,n} : q\in[7], n\in\{2,6,7\}\},\\
\mathcal{I}_{\mathcal{B}_{7}}&=\{v_{q,n} : q\in[7], n\in\{1,3,7\}\}.
\end{align*}
Table \ref{table2.1} lists the intermediate values required by each of the nodes. We cluster each $v_{x,x} : x\in\mathcal{X}$ into $3$-segments
\begin{align*}
v_{1,1} & = \left(v^{\mathcal{B}_{1}}_{1,1},v^{\mathcal{B}_{5}}_{1,1}, 
v^{\mathcal{B}_{7}}_{1,1}\right), 
& v_{2,2} &= \left(v^{\mathcal{B}_{1}}_{2,2}, v^{\mathcal{B}_{2}}_{2,2}, v^{\mathcal{B}_{6}}_{2,2}\right), 
& v_{3,3} &= \left(v^{\mathcal{B}_{2}}_{3,3}, v^{\mathcal{B}_{3}}_{3,3}, v^{\mathcal{B}_{7}}_{3,3}\right), 
& v_{4,4} &= \left(v^{\mathcal{B}_{1}}_{4,4}, v^{\mathcal{B}_{3}}_{4,4}, v^{\mathcal{B}_{4}}_{4,4}\right),\\
v_{5,5} &= \left(v^{\mathcal{B}_{2}}_{5,5}, v^{\mathcal{B}_{4}}_{5,5}, v^{\mathcal{B}_{5}}_{5,5}\right), 
& v_{6,6} & = \left(v^{\mathcal{B}_{3}}_{6,6}, v^{\mathcal{B}_{5}}_{6,6}, v^{\mathcal{B}_{6}}_{6,6}\right), 
& v_{7,7} &= \left(v^{\mathcal{B}_{4}}_{7,7}, v^{\mathcal{B}_{6}}_{7,7},v^{\mathcal{B}_{7}}_{7,7}\right).
\end{align*}
When this is the case, the nodes can collectively send the coded signals listed in Table \ref{table2.2}, with distinct  $\alpha_{1},\alpha_{2},\alpha_{3}\in \mathbb{F}_{2^{\frac{T}{3}}}$. Node $\mathcal{B}_{1}$, for instance, sends the coded signals
\begin{equation}\label{eq:codedsign}
v^{\mathcal{B}_{1}}_{1,1}+ v^{\mathcal{B}_{1}}_{2,2}+ v^{\mathcal{B}_{1}}_{4,4} \mbox{ and } \alpha_{1} v^{\mathcal{B}_{1}}_{1,1} + \alpha_{2} v^{\mathcal{B}_{1}}_{2,2} + \alpha_{3} v^{\mathcal{B}_{1}}_{4,4}.
\end{equation}

After receiving the signals in \eqref{eq:codedsign}, node $\mathcal{B}_{2}$ can individually decode the intermediate values $v^{\mathcal{B}_{1}}_{1,1}$ by using the locally computed intermediate value $v^{\mathcal{B}_{1}}_{2,2}$. Similarly, node $\mathcal{B}_{2}$ can decode the required intermediate values $v^{\mathcal{B}_{5}}_{1,1}$ and $v^{\mathcal{B}_{7}}_{1,1}$ from nodes $\mathcal{B}_{5}$ and $\mathcal{B}_{7}$, respectively. Doing so allows node $\mathcal{B}_{2}$ to decode $v_{1,1}$. It is straightforward to verify that the situation holds for each node and the required value $v_{x,x} : x \in[7]$.

Let us now consider $v_{x,y} : x \neq y$. Node $\mathcal{B}_{1}$, for example, sends the coded signal $v_{1,2} + v_{1,4}$. Upon receiving the signal, nodes $\mathcal{B}_{3}$ and $\mathcal{B}_{4}$ can individually decode $v_{1,2}$ by using the locally computable $v_{1,4}$. Nodes $\mathcal{B}_{2}$ and $\mathcal{B}_{6}$ can individually decode $v_{1,4}$ from the locally computable $v_{1,2}$. Similarly, all other nodes can obtain their respective intermediate values. Thus, the communication load of our scheme is $L=\frac{7 \cdot \frac{2}{3} + 3 \cdot 7}{7 \cdot 7} = \frac{11}{21}$. When $K=7$, $r=3$, and $s=4$, we reproduce a cascaded CDC scheme from \cite{jiang2022} with $N=Q=7$ whose communication load $L'=\frac{7-3}{7-1}=\frac{2}{3}$ is larger than that of ours.

\begin{table}
\caption{Coded signals sent by the nodes in $\mathfrak{B}$}
\label{table2.2}
\renewcommand{\arraystretch}{1.3}
\centering
\begin{tabular}{cccc}
\toprule
$\mathcal{B}_{1}$ & $\mathcal{B}_{2}$ & $\mathcal{B}_{3}$ & $\mathcal{B}_{4}$  \\
\midrule
$v^{\mathcal{B}_{1}}_{1,1}+v^{\mathcal{B}_{1}}_{2,2}+v^{\mathcal{B}_{1}}_{4,4}$ & $v^{\mathcal{B}_{2}}_{2,2}+v^{\mathcal{B}_{2}}_{3,3}+v^{\mathcal{B}_{2}}_{5,5}$ & $v^{\mathcal{B}_{3}}_{3,3}+v^{\mathcal{B}_{3}}_{4,4}+v^{\mathcal{B}_{3}}_{6,6}$ & $v^{\mathcal{B}_{4}}_{4,4}+v^{\mathcal{B}_{4}}_{5,5}+v^{\mathcal{B}_{4}}_{7,7}$ \\
$\alpha_{1}v^{\mathcal{B}_{1}}_{1,1}+\alpha_{2}v^{\mathcal{B}_{1}}_{2,2}+\alpha_{3}v^{\mathcal{B}_{1}}_{4,4}$ & $\alpha_{1}v^{\mathcal{B}_{2}}_{2,2}+\alpha_{2}v^{\mathcal{B}_{2}}_{3,3}+\alpha_{3}v^{\mathcal{B}_{2}}_{5,5}$ & $\alpha_{1}v^{\mathcal{B}_{3}}_{3,3}+\alpha_{2}v^{\mathcal{B}_{3}}_{4,4}+\alpha_{3}v^{\mathcal{B}_{3}}_{6,6}$ & $\alpha_{1}v^{\mathcal{B}_{4}}_{4,4}+\alpha_{2}v^{\mathcal{B}_{4}}_{5,5}+\alpha_{3}v^{\mathcal{B}_{4}}_{7,7}$ \\
$v_{1,2}+v_{1,4}$ & $v_{2,3}+v_{2,5}$ & $v_{3,4}+v_{3,6}$ & $v_{4,5}+v_{4,7}$  \\
$v_{2,4}+v_{2,1}$ & $v_{3,5}+v_{3,2}$ & $v_{4,6}+v_{4,3}$ & $v_{5,4}+v_{5,7}$  \\
$v_{4,1}+v_{4,2}$ & $v_{5,2}+v_{5,3}$ & $v_{6,3}+v_{6,4}$ & $v_{7,4}+v_{7,5}$  \\
\midrule
\midrule
$\mathcal{B}_{5}$ & $\mathcal{B}_{6}$ & $\mathcal{B}_{7}$ &\\
\midrule
$v^{\mathcal{B}_{5}}_{1,1}+v^{\mathcal{B}_{5}}_{5,5}+ v^{\mathcal{B}_{5}}_{6,6}$ & $v^{\mathcal{B}_{6}}_{2,2} + v^{\mathcal{B}_{6}}_{6,6} + v^{\mathcal{B}_{6}}_{7,7}$ & $v^{\mathcal{B}_{7}}_{1,1} + v^{\mathcal{B}_{7}}_{3,3}+v^{\mathcal{B}_{7}}_{7,7}$ & \\
$\alpha_{1}v^{\mathcal{B}_{5}}_{1,1}+\alpha_{2}v^{\mathcal{B}_{5}}_{5,5}+\alpha_{3}v^{\mathcal{B}_{5}}_{6,6}$ & $\alpha_{1}v^{\mathcal{B}_{6}}_{2,2}+\alpha_{2}v^{\mathcal{B}_{6}}_{6,6}+\alpha_{3}v^{\mathcal{B}_{6}}_{7,7}$ & $\alpha_{1}v^{\mathcal{B}_{7}}_{1,1}+\alpha_{2}v^{\mathcal{B}_{7}}_{3,3}+\alpha_{3}v^{\mathcal{B}_{7}}_{7,7}$ &\\
$v_{1,5}+v_{1,6}$ & $v_{2,6}+v_{2,7}$ & $v_{1,3}+v_{1,7}$ &\\
$v_{6,1}+v_{6,5}$ & $v_{6,2}+v_{6,7}$ & $v_{3,1}+v_{3,7}$ &\\
$v_{5,1}+v_{5,6}$ & $v_{7,2}+v_{7,6}$ & $v_{7,1}+v_{7,3}$&\\
\bottomrule
\end{tabular}
\end{table}
\end{example}

\subsection{Construction Two}\label{subsec:scheme2}
Cheng, Wu, and Li in \cite{cheng2023} constructed some asymptotically optimal cascaded CDC schemes by using $t$-designs and $t$-GDDs with $t \geq 2$. We propose a construction of such schemes based on $1$-designs. For the case of $r=s$ we use almost difference (AD) sets. For any $(n,k,\lambda,\mu)$ AD set $(A,D)$ with $\lambda < k-1$, we denote by $(A,+)$ the abelian group $\{0,1,\ldots,n-1\}$ under addition and by $D$ the set $\{i_{1},i_{2},\ldots,i_{k} : i_{t}\in\{0,1,\ldots,n-1\} \mbox{ for any } t\in [k] \}$. There exist $n$ subsets $\mathcal{B}_{r}=\{i'_{1},i'_{2},\cdots,i'_{k}\} \subseteq A$, where $i'_{t} \equiv i_{t}+r-1 \pmod{n}$ with $t \in[k]$ and $r\in [n]$. By the definition of {\tt difference} function, when $\lambda<k-1$, we know that $\mathcal{B}_{u} \neq \mathcal{B}_{v}$ for any $u,v\in [n]$ such that $m\neq n$. Letting $\mathcal{B} = \{\mathcal{B}_{1},\mathcal{B}_{2},\ldots,\mathcal{B}_{n}\}$, we confirm that $(A,\mathfrak{B})$ is a $1$-design with parameters $(n,k,k)$. To verify that $(A,\mathfrak{B})$ is not a $2$-design, we observe that, if $\{a,b\} \subseteq A$ with $|{\rm diff}_{D}(a-b)| = \lambda+1$, then $\{a,b\}$ is contained in the $\lambda+1$ elements of $\mathfrak{B}$. If $\{c,d\}\subseteq A$ with $|{\rm diff}_{D}(c-d)|=\lambda$, then $\{c,d\}$ is contained in the $\lambda$ elements of $\mathfrak{B}$. Focusing on the $\binom{A}{2}$ subsets of two elements in $A$, there are, respectively, $\frac{nt}{2}$ and $\frac{(n-1-t)n}{2}$ such subsets which are contained in $\lambda$ and $\lambda+1$ elements of $\mathfrak{B}$.

\begin{example} \label{ex1.1}
We use $A=\{0,1,2,3,4,5\}$ to form an abelian group under addition. We verify that $D=\{0,1,3\}$ is a $(6,3,1,4)$ AD set, where the function ${\rm diff}_{D}(x)$ takes on $1$, in total, $4$ times, if $x\in\{1,2,4,5\}$, and takes on $2$ once if $x=3$. Our construction yields the composite structure $(A,\mathfrak{B})$, where
\[
\mathfrak{B} = \{\{0,1,3\},\{1,2,4\},\{2,3,5\},\{3,4,0\},\{4,5,1\},\{5,2,0\}\}.
\]
We confirm that $(A,\mathfrak{B})$ is a $1$-design with parameters $(6,3,3)$. It, however, is not a $2$-design since the pairs 
\[
\{0,3\}, \{1,4\}, \{2,5\}
\]
are contained in $2$ elements of $\mathfrak{B}$, but the pairs 
\[
\{0,1\}, \{0,2\}, \{0,4\}, \{0,5\}, \{1,2\}, \{1,3\}, \{1,5\}, \{2,3\}, \{2,4\}, \{3,4\}, \{3,5\}, \{4,5\}
\]
are contained in only a single element of $\mathfrak{B}$.
\end{example}

\begin{example}\label{ex1.2}
Let $(A,+)=\{0,1,2,3,4,5\}$ be the abelian group. We confirm that $D=\{0,1\}$ is a $(6,2,0,3)$ AD set. Its ${\rm diff}_{D}(x)$ takes on $1$, in total, twice for $x \in\{1,5\}$ and $0$, in total, $3$ times if $x\in\{2,3,4\}$. We have the composite structure $(A,\mathfrak{B})$ with
\[
\mathfrak{B} = \{\{0,1\},\{1,2\},\{2,3\},\{3,4\},\{4,5\},\{5,0\}\}.
\]
We verify that $(A,\mathfrak{B})$ is a $1$-design with parameter $(6,2,2)$. It is not a $2$-design since the pairs 
\[
\{0,1\}, \{1,2\}, \{2,3\}, \{3,4\}, \{4,5\}, \{0,5\}
\]
are contained in a single element of $\mathfrak{B}$, but the pairs 
\[
\{0,2\}, \{0,3\}, \{0,4\}, \{1,3\}, \{1,4\}, \{1,5\}, \{2,4\}, \{2,5\}, \{3,5\}
\]
are not contained in any element of $\mathfrak{B}$.
\end{example}

We refine our construction into two cases: $\lambda\geq1$ and $\lambda=0$. We start with the case of $\lambda\geq1$ and construct a CDC scheme with $N$ nodes, $\mathcal{K}=\mathcal{B}$, $n$ files in $\mathcal{W} = \{w_{0},w_{1},\ldots,w_{n-1}\}$, and $n$ functions in $\mathcal{Q} = \{\phi_{0},\phi_{1},\ldots,\phi_{n-1}\}$. Each node stores $k$ files and each {\tt Reduce} function is computed by $s$ nodes.

During the {\tt Map} phase, let each node $\mathcal{B}\in \mathfrak{B}$ store the files in $Z_{\mathcal{B}}=\{w_{x} : x\in\mathcal{B}, \ x\in\mathcal{X}\}$. Since the cardinality of any block is $|\mathcal{B}| = k$, the computation load is
\[
r=\frac{\sum^{n}_{i=1}\mid Z_{i}\mid}{n} = \frac{kn}{n}=k.
\]
In the {\tt Shuffle} phase, let each node $\mathcal{B} \in \mathfrak{B}$ be arranged to compute the {\tt Reduce} functions in
\[
\mathcal{Q}_{\mathcal{B}} = \left\{u_{y}=\phi_{y}(w_{0},w_{1},\ldots,w_{n-1}) : y\in A, y\in \mathcal{B}\right\}
\]
Using the stored files and the functions in $\mathcal{Q}$, node $\mathcal{B}$ can compute the intermediate values 
\[
\mathcal{I}_{\mathcal{B}} = \left\{v_{y,x} = g_{y,x}(w_{x}) : x,y\in A,x\in \mathcal{B}\right\}.
\] 
For any $x,y\in A$ and any block $\mathcal{B}\in \mathfrak{B}$, the intermediate value $v_{y,x}$ is required. It is \emph{not} locally computable by node $\mathcal{B}$ if and only if $y\in \mathcal{B}$ and $x\notin \mathcal{B}$. On the other hand, $v_{y,x}$ is locally computable by node $\mathcal{B}$ if and only if $x\in \mathcal{B}$. We devise our delivery strategy accordingly.

If $x,y\in A$ are such that $|{\rm diff}_{D}(x-y)| =\lambda+1$, then there exist $\lambda+1$ nodes with access to the pair of files $(x,y)$. We call these nodes $\mathcal{B}_{1,j}$ with $j \in[\lambda+1]$. There are $k-\lambda-1$ nodes with access to file $x$ but not file $y$. We label these nodes $\mathcal{B}_{2,u}$ with $u \in[k-\lambda-1]$. There are $k-\lambda-1$ nodes with access to file $y$ but not file $x$. We name these nodes $\mathcal{B}_{3,v}$ with $v\in[k-\lambda-1]$. 

Our delivery strategy must allow for the exchange of relevant intermediate values among the nodes. Each node $\mathcal{B}_{1,j}$ can locally compute $v_{x,y}$ and $v_{y,x}$ since it stores files $w_{x}$ and $w_{y}$. Each node $\mathcal{B}_{2,u}$ can locally compute $v_{y,x}$ since it stores file $w_{x}$ but requires $v_{x,y}$ from some other nodes. This node does not store $w_{y}$ but is assigned to compute the {\tt Reduce} function $u_{y}$. Each node $\mathcal{B}_{3,v}$ can locally compute $v_{x,y}$ since it stores file $w_{y}$ but requires $v_{y,x}$ from some other nodes. This node does not store $w_{x}$ but is assigned to compute the {\tt Reduce} function $u_{x}$.

We divide $v_{x,y}$ and $v_{y,x}$ into $\lambda+1$ sub-intermediate values
\[
v_{y,x}=\left\{v^{(1)}_{y,x},v^{(2)}_{y,x},\ldots,v^{(\lambda+1)}_{y,x}\right\} \mbox{ and } v_{x,y}=\left\{v^{(1)}_{x,y},v^{(2)}_{x,y},\ldots,v^{(\lambda+1)}_{x,y}\right\}.
\]
Node $\mathcal{B}_{1,j}$ multicasts  $\left\{v^{(i_{1})}_{y,x} + v^{(i_{1})}_{x,y} : i_{1}\in [\lambda+1] \right\}$ to nodes $\mathcal{B}_{2,i_{2}}$ and $\mathcal{B}_{3,i_{3}}$, with $i_{2},i_{3}\in [k-\lambda-1]$. Hence, any $\mathcal{B}_{2,i_{2}}$ and $\mathcal{B}_{3,i_{3}}$ can derive $v^{(i_{1})}_{x,y}$ and $v^{(i_{1})}_{y,x}$, respectively. Since $(A,D)$ is an $(n,k,\lambda,\mu)$ AD set, for each node $\mathcal{B}\in \mathfrak{B}$ there exist $n-1-\mu$ pairs $(x,y)$, with $\{x,y\} \subseteq \mathcal{B}$, such that $|{\rm diff}_{D}(x-y)| = \lambda+1$. Paying closer attention to the $\binom{A}{2}$ subsets of two elements in $A$, we infer that there are $\frac{(n-1-\mu)n}{2}$ such subsets which are contained in $\lambda+1$ elements of $\mathfrak{B}$. Thus, in this particular delivery strategy, there are exactly $\frac{(n-1-\mu)(\lambda+1)n}{2}$ transmitted sub-intermediate values, each of which has $\frac{T}{\lambda+1}$ bits. The total number of bits transmitted by the nodes is $\frac{(n-1-\mu)Tn}{2}$. 

If $u,v\in A$ are such that $|{\rm diff}_{D}(u-v)| = \lambda$, then we have the desired exchange scheme. In total, the number of bits transmitted is $\frac{\mu Tn}{2}$ for $\frac{\mu Tn+(n-1-\mu)\,Tn}{2}=\frac{n(n-1)T}{2}$ signals. The communication load is $L=\frac{n(n-1)T}{2n^{2}T}=\frac{n-1}{2n}$, leading us to the following theorem.

\begin{thm} \label{thm2.1}
Given an $(n,k,\lambda,\mu)$ almost different set $(A,D)$ with $1\leq\lambda<k-1$, one can construct a CDC scheme with $n$ distributed computing nodes, $N=n$ files, and $Q=n$ output functions such that each output function is computed by $s=k$ nodes. The scheme's respective computation and communication loads are  $r=k$ and $L=\frac{n-1}{2n}$.
\end{thm}

\begin{example}
Continuing from Example \ref{ex1.1}, we can construct the following coded distributed computing. When $N=Q=K=6$, we have $6$ files $\mathcal{W}=\{w_{0},w_{1},\cdots,w_{5}\}$ and $6$ output functions $\mathcal{Q}=\{\phi_{0},\phi_{1},\ldots,\phi_{5}\}$. In the {\tt Map} phase, the nodes and their respective stored files are 
\begin{align*}
\mathcal{Z}_{\mathcal{B}_{1}}&=\{w_{0},w_{1},w_{3}\}, &\mathcal{Z}_{\mathcal{B}_{2}}&=\{w_{1},w_{2},w_{4}\}, &\mathcal{Z}_{\mathcal{B}_{3}}&=\{w_{2},w_{3},w_{5}\}, \\
\mathcal{Z}_{\mathcal{B}_{4}}&=\{w_{3},w_{4},w_{0}\},  &\mathcal{Z}_{\mathcal{B}_{5}}&=\{w_{4},w_{5},w_{1}\},  &\mathcal{Z}_{\mathcal{B}_{6}}&=\{w_{5},w_{0},w_{2}\}.
\end{align*}
The computation load is $r=\frac{3 \cdot 6}{6}=3$.

Let the {\tt Reduce} functions be arranged by nodes, such that each function is computed by $s=3$ nodes, as
\begin{align*}
\mathcal{Q}_{\mathcal{B}_{1}} &=\{\phi_{0},\phi_{1},\phi_{3}\}, &\mathcal{Q}_{\mathcal{B}_{2}}&=\{\phi_{1},\phi_{2},\phi_{4}\},  &\mathcal{Q}_{\mathcal{B}_{3}}&=\{\phi_{2},\phi_{3},\phi_{5}\},
\\
\mathcal{Q}_{\mathcal{B}_{4}}&=\{\phi_{3},\phi_{4},\phi_{0}\}, &\mathcal{Q}_{\mathcal{B}_{5}}&=\{\phi_{4},\phi_{5},\phi_{1}\}, &\mathcal{Q}_{\mathcal{B}_{6}}&=\{\phi_{5},\phi_{0},\phi_{2}\}.
\end{align*}

The indicated nodes can then locally compute their respective intermediate values
\begin{align*}
\mathcal{I}_{\mathcal{B}_{1}}&=\{v_{q,n} : q\in\{0,1,2,3,4,5\}, n\in\{0,1,3\}\}, 
&\mathcal{I}_{\mathcal{B}_{2}}&=\{v_{q,n} : q\in\{0,1,2,3,4,5\}, n\in\{1,2,4\}\},\\
\mathcal{I}_{\mathcal{B}_{3}}&=\{v_{q,n} : q\in\{0,1,2,3,4,5\}, n\in\{2,3,5\}\}, 
&\mathcal{I}_{\mathcal{B}_{4}}&=\{v_{q,n} : q\in\{0,1,2,3,4,5\}, n\in\{3,4,0\}\},\\
\mathcal{I}_{\mathcal{B}_{5}}&=\{v_{q,n} : q\in\{0,1,2,3,4,5\}, n\in\{4,5,1\}\}, 
&\mathcal{I}_{\mathcal{B}_{6}}&=\{v_{q,n} : q\in\{0,1,2,3,4,5\}, n\in\{5,0,2\}\}.
\end{align*}

Table \ref{table1.1} lists the required intermediate values in relation to the nodes.

\begin{table}[!htbp]
\caption{Intermediate values $\{v_{q,n}\}$ required by the nodes in $\mathfrak{B}$}
\label{table1.1}
\renewcommand{\arraystretch}{1.1}
\centering
\begin{tabular}{c|cccccc}
\toprule
Parameters & \multicolumn{6}{c}{node set $\mathfrak{B}$} \\
\midrule
 & $\mathcal{B}_{1}$ & $\mathcal{B}_{2}$ & $\mathcal{B}_{3}$ & $\mathcal{B}_{4}$ & $\mathcal{B}_{5}$ & $\mathcal{B}_{6}$  \\
\midrule
$q$ & $0,1,3$ & $1,2,4$ & $2,3,5$ & $3,4,0$ & $4,5,1$ & $5,0,2$  \\
\midrule
$n$ & $2,4,5$ & $0,3,5$ & $0,1,4$ & $1,2,5$ & $0,2,3$ & $1,3,4$  \\
\bottomrule
\end{tabular}
\end{table}

We cluster all intermediate values $v_{0,3}$, $v_{3,0}$, $v_{1,4}$, $v_{4,1}$, $v_{5,2}$ and $v_{2,5}$ into the $2$-segments
\begin{align*}
v_{0,3} &= \left(v^{(1)}_{0,3},v^{(2)}_{0,3}\right), 
& v_{1,4} &= \left(v^{(1)}_{1,4},v^{(2)}_{1,4}\right), 
& v_{2,5} &= \left(v^{(1)}_{2,5},v^{(2)}_{2,5}\right),\\
v_{3,0} &= \left(v^{(1)}_{3,0},v^{(2)}_{3,0}\right), 
&v_{4,1} &= \left(v^{(1)}_{4,1},v^{(2)}_{4,1}\right),  
&v_{5,2} &= \left(v^{(1)}_{5,2},v^{(2)}_{5,2}\right).
\end{align*}
The nodes send the coded signals listed in Table \ref{table1.2}. Node $\mathcal{B}_{1}$, for instance, sends  $v_{0,1}+v_{1,0}$. After receiving it, nodes $\mathcal{B}_{4}$ and $\mathcal{B}_{6}$ get the required $v_{0,1}$ since they can locally compute $v_{1,0}$. Similarly, nodes $\mathcal{B}_{2}$ and $\mathcal{B}_{5}$ can obtain $v_{1,0}$ after receiving $v_{0,1}+v_{1,0}$. Inspecting the rest of the nodes, we have analogous situation. The communication load is $\frac{(2+\frac{1}{2})6}{6\cdot 6} = \frac{5}{12}$.
\begin{table}
\caption{Coded signals sent by the nodes in $\mathfrak{B}$}
\label{table1.2}
\centering
\renewcommand{\arraystretch}{1.2}
\begin{tabular}{cccccc}
\toprule
$\mathcal{B}_{1}$ & $\mathcal{B}_{2}$ & $\mathcal{B}_{3}$ & $\mathcal{B}_{4}$ & $\mathcal{B}_{5}$ & $\mathcal{B}_{6}$   \\
\midrule
$v_{0,1}+v_{1,0}$ & $v_{1,2}+v_{2,1}$ & $v_{2,3}+v_{3,2}$ & $v_{4,0}+v_{0,4}$ & $v_{1,5}+v_{5,1}$ & $v_{5,0}+v_{0,5}$   \\
$v^{(1)}_{0,3}+v^{(1)}_{3,0}$ & $v^{(1)}_{1,4}+v^{(1)}_{4,1}$ & $v^{(1)}_{2,5}+v^{(1)}_{5,2}$ & $v^{(2)}_{0,3}+v^{(2)}_{3,0}$ & $v^{(2)}_{1,4}+v^{(2)}_{4,1}$ & $v^{(2)}_{2,5}+v^{(2)}_{5,2}$  \\
$v_{3,1}+v_{1,3}$ & $v_{2,4}+v_{4,2}$ & $v_{3,5}+v_{5,3}$ & $v_{3,4}+v_{4,3}$ & $v_{4,5}+v_{5,4}$ & $v_{0,2}+v_{2,0}$  \\
\bottomrule
\end{tabular}
\end{table}
\end{example}

We continue to the case of $\lambda=0$. If $\{u,w\}\subseteq A$ such that $|{\rm diff}_{D}(u-w)|=1$, then there exists an element of $\mathfrak{B}$ that contain $u$ and $w$. If $\{u,w\}\subseteq A$ such that $|{\rm diff}_{D}(u-w)|=0$, then there is no element of $\mathfrak{B}$ that contains $u$ and $w$. We recall that an $(n,k,0,\mu)$ almost difference sets are also known as \emph{modular Golomb rulers}.

Since the {\tt Map} phase is the same as in the case of $\lambda\geq 1$ above, the computation load is also $r=k$.

In the {\tt Shuffle} phase, if each node $\mathcal{B}\in \mathfrak{B}$ is to compute the {\tt Reduce} functions in
\[
\mathcal{Q}_{\mathcal{B}} = \left\{u_{y}=\phi_{y}(w_{0},w_{1},\ldots,w_{n-1}) : y\in A, y\in \mathcal{B}\right\},
\]
then $s=r=k$. Using the stored files and the functions in $\mathcal{Q}$, node $\mathcal{B}$ can compute the intermediate values 
\[
\mathcal{I}_{\mathcal{B}} = \{v_{y,x} = g_{y,x}(w_{x}) : x,y\in A,x\in \mathcal{B}\}.
\]
Hence, for any $x,y\in A$ and any block $\mathcal{B}\in \mathfrak{B}$, the intermediate value $v_{y,x}$ is required but not locally computable by node $\mathcal{B}$ if and only if $y \in \mathcal{B}$ and $x \notin \mathcal{B}$. It is locally computable by node $\mathcal{B}$ if and only if $x\in \mathcal{B}$. By a similar analysis as in the case of $\lambda\geq 1$, there exist $\frac{k(k-1)n}{2}$ pairs of elements of $A$ which are contained by elements of $\mathfrak{B}$. On the other hand, there exist $\frac{n\mu}{2}$ pairs of elements of $A$ which are not contained in any element of $\mathfrak{B}$. We adjust the delivery strategy accordingly. 

First, for the $\frac{k(k-1)n}{2}$ pairs, we use the delivery strategy in the proof of Theorem \ref{thm2.1}. There are $\frac{k(k-1)nT}{2}$ transmitted signals in total. Second, for the remaining $\frac{n\mu}{2}$ intermediate values $v_{u,w}$ for which $\{u,w\}$ is not contained in any element of $\mathfrak{B}$, no node can broadcast the coded signal $v_{u,w}+v_{w,u}$. If $u,w \in A$ are files such that $|{\rm diff}_{D}(u-w)|=0$, then $u$ and $w$ are stored by nodes whose block representatives contain $u$. Both files are required by nodes whose block representatives contain $w$. We collect the respective blocks containing $u$ and $w$ into sets
\[
\mathfrak{B}_{u} =\{\mathcal{B}^{u}_{1}, \mathcal{B}^{u}_{2},\ldots,\mathcal{B}^{u}_{k}\} \mbox{ and } \mathfrak{B}_{w} = \{\mathcal{B}^{w}_{1},\mathcal{B}^{w}_{2},\ldots, \mathcal{B}^{w}_{k}\}.
\]
We split $v_{u,w}$ into $k$ sub-intermediate values 
\[
v^{(1)}_{u,w}, v^{(2)}_{u,w}, \ldots, v^{(k)}_{u,w}.
\]
Node $\mathcal{B}^{w}_{i}$ sends $v^{(i)}_{u,w}$ to nodes in $\mathfrak{B}_{u}$. Clearly, each node in $\mathfrak{B}_u$ can obtain $v_{u,w}$ from all $k$ sub-intermediate values sent by the nodes in $\mathfrak{B}_{w}$. In total, there are $(n\mu+\frac{k(k-1)n}{2})T$ transmitted signals. Since $k(k-1)=n-1-\mu$, the system transmits $(n(n-1)-\frac{k(k-1)n}{2})T$ signals. From the above discussion, the communication load is $L=\frac{2n-2-k(k-1)}{2n}$. We have thus proved the next result.
\begin{thm} \label{thm2.2}
Given an $(n,k,0,\mu)$ almost different set $(A,D)$, one can construct a CDC scheme with $n$ distributed computing nodes, $N=n$ files, and $Q=n$ output functions such that each output function is computed by $s=k$ nodes. The respective computation and communication loads are $r=k$ and $L=\frac{2n-2-k(k-1)}{2n}$.
\end{thm}

\begin{example}
Continuing from Example \ref{ex1.2}, we construct the following CDC scheme. When $N=Q=K=6$, we have $6$ files in $\mathcal{W}=\{w_{0},w_{1},\ldots,w_{5}\}$ and $6$ output functions in $\mathcal{Q}=\{\phi_{0},\phi_{1},\ldots,\phi_{5}\}$. In the {\tt Map} phase, the nodes and their respective stored files are
\begin{align*}
\mathcal{Z}_{\mathcal{B}_{1}} &=\{w_{0},w_{1}\},  
&\mathcal{Z}_{\mathcal{B}_{2}}&=\{w_{1},w_{2}\},  &\mathcal{Z}_{\mathcal{B}_{3}}&=\{w_{2},w_{3}\},\\
\mathcal{Z}_{\mathcal{B}_{4}}& =\{w_{3},w_{4}\},  &\mathcal{Z}_{\mathcal{B}_{5}}&=\{w_{4},w_{5}\},  &\mathcal{Z}_{\mathcal{B}_{6}}&=\{w_{5},w_{0}\}.
\end{align*}
Hence, the computation load is $r=\frac{2 \cdot 6}{6}=2$.

Let the {\tt Reduce} functions be arranged by nodes such that each function is computed by $s=2$ nodes as 
\begin{align*}
\mathcal{Q}_{\mathcal{B}_{1}}&=\{\phi_{0},\phi_{1}\},  
&\mathcal{Q}_{\mathcal{B}_{2}}&=\{\phi_{1},\phi_{2}\},   &\mathcal{Q}_{\mathcal{B}_{3}}&=\{\phi_{2},\phi_{3}\},\\
\mathcal{Q}_{\mathcal{B}_{4}}&=\{\phi_{3},\phi_{4}\},  &\mathcal{Q}_{\mathcal{B}_{5}}&=\{\phi_{4},\phi_{5}\},  &\mathcal{Q}_{\mathcal{B}_{6}}&=\{\phi_{5},\phi_{0}\}.
\end{align*}
The indicated nodes can then compute the respective intermediate values
\begin{align*}
\mathcal{I}_{\mathcal{B}_{1}} &=\{v_{q,n} : q\in\{0,1,2,3,4,5\}, n\in\{0,1\}\}, 
&\mathcal{I}_{\mathcal{B}_{2}}&=\{v_{q,n} : q\in\{0,1,2,3,4,5\}, n\in\{1,2\}\},\\
\mathcal{I}_{\mathcal{B}_{3}}&=\{v_{q,n} : q\in\{0,1,2,3,4,5\}, n\in\{2,3\}\}, 
&\mathcal{I}_{\mathcal{B}_{4}}&=\{v_{q,n} : q\in\{0,1,2,3,4,5\}, n\in\{3,4\}\},\\
\mathcal{I}_{\mathcal{B}_{5}}&=\{v_{q,n} : q\in\{0,1,2,3,4,5\}, n\in\{4,5\}\}, 
&\mathcal{I}_{\mathcal{B}_{6}}&= \{v_{q,n} : q\in\{0,1,2,3,4,5\}, n\in\{5,0\}\}.
\end{align*}
Table \ref{table1.3} lists the required intermediate values in relation to the nodes. 

\begin{table}
	\caption{Intermediate values $\{v_{q,n}\}$ required by nodes in $\mathfrak{B}$}
	\label{table1.3}
	\renewcommand{\arraystretch}{1.2}
	\centering
	\begin{tabular}{c|cccccc}
		\toprule
		Parameters & \multicolumn{6}{c}{node Set $\mathfrak{B}$} \\
		\midrule
		$q,n$ & $\mathcal{B}_{1}$ & $\mathcal{B}_{2}$ & $\mathcal{B}_{3}$ & $\mathcal{B}_{4}$ & $\mathcal{B}_{5}$ & $\mathcal{B}_{6}$  \\
		\midrule
		$q$ & $0,1$ & $1,2$ & $2,3$ & $3,4$ & $4,5$ & $5,0$  \\
		\midrule
		$n$ & $2,3,4,5$ & $0,3,4,5$ & $0,1,4,5$ & $0,1,2,5$ & $0,1,2,3$ & $1,2,3,4$  \\
		\bottomrule
	\end{tabular}
\end{table}

We cluster the intermediate values 
\[
v_{0,2}, v_{0,3}, v_{0,4}, v_{1,3}, v_{1,4}, v_{1,5}, v_{2,4}, v_{2,5}, v_{2,0}, v_{3,1}, v_{3,5}, v_{3,0}, v_{4,1}, v_{4,2}, v_{4,0}, v_{5,2}, v_{5,3}, v_{5,1}
\]
into the $2$-segments
\begin{align*}
v_{0,2} &= \left(v^{(1)}_{0,2},v^{(2)}_{0,2}\right), & v_{0,3} &= \left(v^{(1)}_{0,3},v^{(2)}_{0,3}\right), & v_{0,4} &= \left(v^{(1)}_{0,4},v^{(2)}_{0,4}\right),\\
v_{1,3} &= \left(v^{(1)}_{1,3},v^{(2)}_{1,3}\right), 
& v_{1,4} &=\left(v^{(1)}_{1,4},v^{(2)}_{1,4}\right), 
& v_{1,5} &=\left(v^{(1)}_{1,5},v^{(2)}_{1,5}\right),\\
v_{2,4}&=\left(v^{(1)}_{2,4},v^{(2)}_{2,4}\right), &v_{2,5}&=\left(v^{(1)}_{2,5},v^{(2)}_{2,5}\right),  &v_{2,0}&=\left(v^{(1)}_{2,0},v^{(2)}_{2,0}\right), \\
v_{3,5}&=\left(v^{(1)}_{3,5},v^{(2)}_{3,5}\right),  &v_{3,0}&=\left(v^{(1)}_{3,0},v^{(2)}_{3,0}\right),  &v_{3,1}&=\left(v^{(1)}_{3,1},v^{(2)}_{3,1}\right), \\
v_{4,2}&=\left(v^{(1)}_{4,2},v^{(2)}_{4,2}\right),  &v_{4,0}&=\left(v^{(1)}_{4,0},v^{(2)}_{4,0}\right),  &v_{4,1}&=\left(v^{(1)}_{4,1},v^{(2)}_{4,1}\right), \\
v_{5,2}&=\left(v^{(1)}_{5,2},v^{(2)}_{5,2}\right),  &v_{5,3}&=\left(v^{(1)}_{5,3},v^{(2)}_{5,3}\right),  &v_{5,1}&=\left(v^{(1)}_{5,1},v^{(2)}_{5,1}\right).
\end{align*}
In this case, the nodes can send the coded signals listed in Table \ref{table1.4}. Node $\mathcal{B}_{1}$, for example, sends $v_{0,1}+v_{1,0}$. After receiving the signal, nodes $\mathcal{B}_{4}$ and $\mathcal{B}_{6}$ can obtain the intermediate value $v_{0,1}$ because they can locally compute $v_{1,0}$. Similarly, nodes $\mathcal{B}_{2}$ and $\mathcal{B}_{5}$ can obtain the required $v_{1,0}$ after receiving $v_{0,1}+v_{1,0}$. On the other hand, nodes $\mathcal{B}_{1}$ and $\mathcal{B}_{6}$ send the respective coded signals $v^{(1)}_{0,2}$ and $v^{(2)}_{0,2}$. Upon receiving $v^{(1)}_{0,2}$ and $v^{(2)}_{0,2}$, nodes $\mathcal{B}_{2}$ and $\mathcal{B}_{3}$ can obtain $v_{0,1}$. The rest of the nodes can obtain their respective required intermediate values in a similar manner. The communication load is $ \frac{(1+ 6 \cdot \frac{1}{2}) 6}{6 \cdot 6}=\frac{2}{3}$.

\begin{table}
\caption{Coded signals sent by nodes in $\mathfrak{B}$}
\label{table1.4}
\renewcommand{\arraystretch}{1.4}
\centering
\begin{tabular}{cccccc}
\toprule
$\mathcal{B}_{1}$ & $\mathcal{B}_{2}$ & $\mathcal{B}_{3}$ & $\mathcal{B}_{4}$ & $\mathcal{B}_{5}$ & $\mathcal{B}_{6}$   \\
\midrule
$v_{0,1}+v_{1,0}$ & $v_{1,2}+v_{2,1}$ & $v_{2,3}+v_{3,2}$ & $v_{4,0}+v_{0,4}$ & $v_{1,5}+v_{5,1}$ & $v_{5,0}+v_{0,5}$   \\
\midrule
$v^{(1)}_{0,2}$ & $v^{(1)}_{1,3}$ & $v^{(1)}_{2,4}$ & $v^{(1)}_{3,5}$ & $v^{(1)}_{4,0}$ & $v^{(1)}_{5,1}$ \\
$v^{(1)}_{0,3}$ & $v^{(1)}_{1,4}$ & $v^{(1)}_{2,5}$ & $v^{(1)}_{3,0}$ & $v^{(1)}_{4,1}$ & $v^{(1)}_{5,2}$ \\
$v^{(1)}_{0,4}$ & $v^{(1)}_{1,5}$ & $v^{(1)}_{2,0}$ & $v^{(1)}_{3,1}$ & $v^{(1)}_{4,2}$ & $v^{(1)}_{5,3}$ \\
$v^{(2)}_{1,3}$ & $v^{(2)}_{2,4}$ & $v^{(2)}_{3,5}$ & $v^{(2)}_{4,0}$ & $v^{(2)}_{5,1}$ & $v^{(2)}_{0,2}$ \\
$v^{(2)}_{1,4}$ & $v^{(2)}_{2,5}$ & $v^{(2)}_{3,0}$ & $v^{(2)}_{4,1}$ & $v^{(2)}_{5,2}$ & $v^{(2)}_{0,3}$ \\
$v^{(2)}_{1,5}$ & $v^{(2)}_{2,0}$ & $v^{(2)}_{3,1}$ & $v^{(2)}_{4,2}$ & $v^{(2)}_{5,3}$ & $v^{(2)}_{0,4}$ \\
\bottomrule
\end{tabular}
\end{table}
\end{example}
\begin{remark}
Although, the schemes in Theorems \ref{thm2.1} and \ref{thm2.2} are quite similar to the scheme in \cite{cheng2023}, we can obtain an asymptotically optimal cascaded CDC scheme with different parameters. The next section focuses on their performance for comparative purposes.
\end{remark}

\section{Performance Comparison and Concluding Remarks}\label{sec:perform}

For fixed $(r,s)$, the number of files $N=\binom{K}{r}$ and functions $Q=\binom{K}{s}$ in the Li-CDC schemes grow fast as the number of computing nodes $K$ increases. In practical scenarios, as Konstantinidis and Ramamoorthy have shown in \cite{kon2020}, this fast growth is detrimental to the performance of the schemes. The number of input files and output functions in each of our new schemes, in contrast, are \emph{equal to} the number of the computing nodes, confirming the superiority of  our schemes. 

What about the respective communication loads? This section compares the communication loads of our scheme in Subsection \ref{subsec:scheme1} and that of the Li-CDC. Jiang, Wang, and Zhou in \cite{jiang2022} have constructed an asymptotically optimal cascaded CDC scheme with $r\neq s$ based on symmetric designs. Their scheme, which we call \emph{Jiang-CDC} for ease of reference, has larger communication load, given the same input files, output functions, and computation load than our scheme in Subsection \ref{subsec:scheme2}. Here we compare the communication load of our scheme and that of Jiang-CDC in \cite[Theorem 2]{jiang2022}.

\subsection{On the CDC Schemes from Theorem \ref{thm2.3}}
Using a $(v,t,\lambda)$ SD, a Jiang-CDC scheme with $r=t$ and $s=v-t$ has communication load $L_{\rm Jiang}=\frac{v-t}{v-1}$. From the same $(v,t,\lambda)$ SD, Theorem \ref{thm2.3} gives us a CDC scheme with $r=s$ and (minimum) communication load $L_{\rm ours}=\frac{(v-1)^{2}-tv+v}{v(v-1)}$. It is straightforward to prove that $L_{\rm ours}$ is smaller than $L_{\rm Jiang}$ on the same number of input files, output functions, computing nodes, $r$, and $s$. Let a suitable $(v,t,\lambda)$ SD be given. For a contradiction, let us assume that $L_{\rm Jiang} \leq L_{\rm ours}$. Hence, $\frac{v-t}{v-1}\leq  \frac{(v-1)^{2}-tv+v}{v(v-1)}$, which is equivalent to $v\leq 1$. It is then clear that $L_{\rm Jiang} \leq L_{\rm ours}$ if and only if $v\leq 1$, which contradicts the very definition of a symmetric design. 

\begin{figure*}[ht!]
\caption{Communication load comparisons of schemes in Theorem \ref{thm2.3} and \cite[Theorem 2]{jiang2022} based on the specified symmetric designs.}
\centering
\begin{tabular}{cc}
\includegraphics[width=0.45\linewidth]{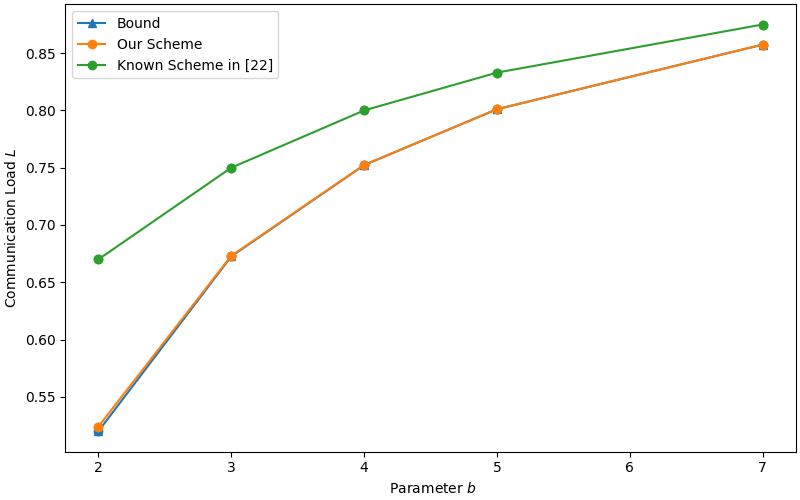} & 
\includegraphics[width=0.45\linewidth]{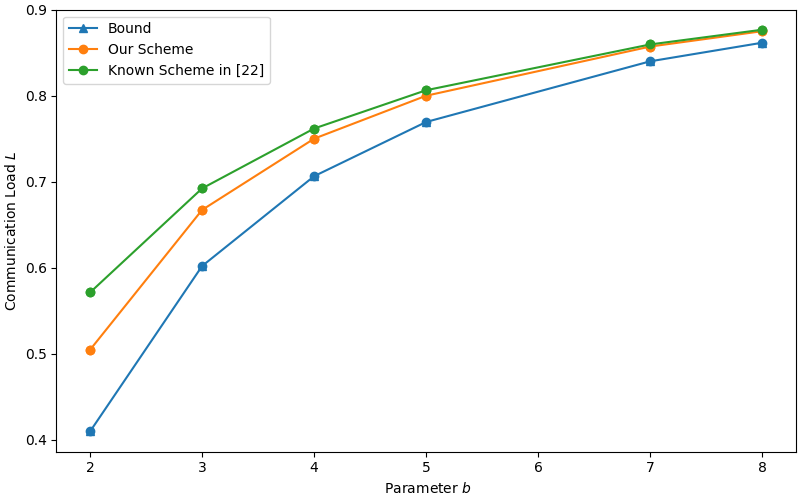}\\
(a) from $(b^{2}+b+1,b+1,1)$ SD & 
(b) from $(b^{3}+b^{2}+b+1,b^{2}+b+1,b+1)$ SD\\
\includegraphics[width=0.45\linewidth]{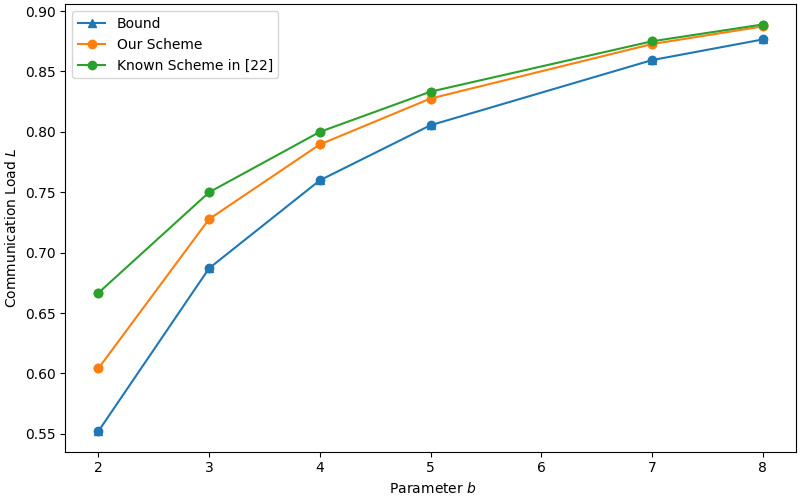} & 
\includegraphics[width=0.45\linewidth]{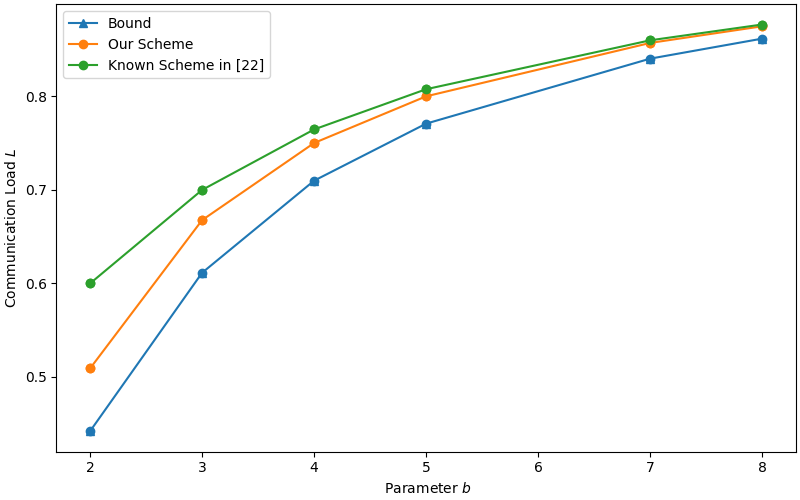}\\
(c) from $(b^{3}+2b^{2},b^{2}+b,b)$ SD & 
(d) from $(b^{3}+b+1,b^{2}+1,b)$ SD\\
\end{tabular}
\label{Figure41}
\end{figure*}

We know that Jiang-CDC schemes constructed from the symmetric designs in Table \ref{table0.1} are all asymptotically optimal. Thus, using the same symmetric designs, our cascaded CDC schemes are also asymptotically optimal. Figure \ref{Figure41} 
provides concrete performance comparison between our CDC schemes and Jiang-CDC schemes based on the specified SDs.
 
\subsection{On the CDC Schemes from Theorem \ref{thm2.2}}
For any prime $p$, Lemma \ref{lem0.2} and Theorem \ref{thm2.1} lead to a construction of a class of CDC scheme. The class has $r=s=p-1$, $K=p^{2}-p$, and communication load $L_{1}:=\frac{p^{2}+p-4}{2(p^{2}-p)}$. The class has $p^{2}-p$ input files and the same number, $p^{2}-p$, of output functions. We establish that schemes in this class are asymptotically optimal, that is, $\frac{L_{1}}{L_{\rm Li}}$ converges to $1$ when $p$ is large. We begin with the following lemma whose proof will be given in the appendix.
\begin{lem} \label{lem3.1}
For a positive integer $p\geq 5$, 
\begin{equation} \label{for31}
\sum^{p-1}_{\ell=0} \ell \binom{p^{2}-2p+1}{\ell} \binom{p-1}{\ell} >(p-3) \, \binom{p^{2}-p}{p-1}.
\end{equation}
\end{lem}

\begin{figure}[ht!]
\centering
\caption{Comparison of communication loads of our scheme and the Li-CDC scheme over $(p^{2}-p,p-1)$ modular Golomb rulers for a range of $p$.}\label{Figure40}
\includegraphics[width=0.7\linewidth]{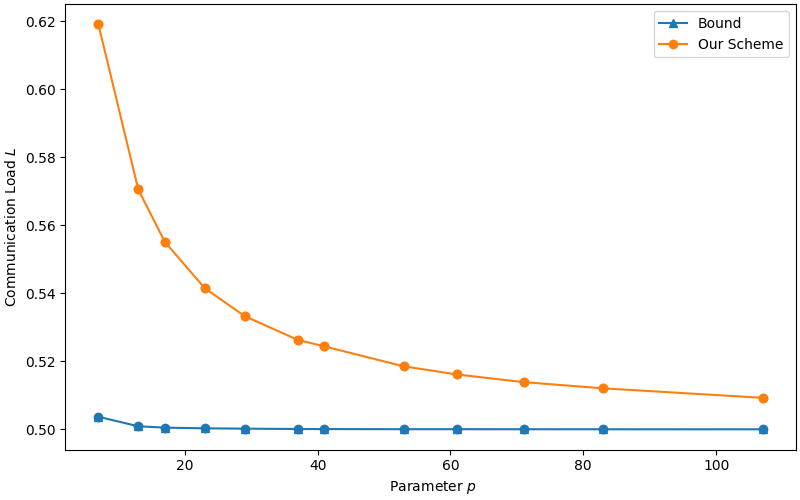}
\end{figure}

If $p$ is large, then $p^{2}-p>2(p-1)$. Taking $r=s=p-1$ and $K=p^{2}-p$ in Lemma \ref{lem0.1} yields 
\[
L_{\rm Li} =\sum^{2(p-1)}_{\ell=p} \left(\frac{\ell-(p-1)}{\ell-1}
\,
\frac{\binom{p^{2}-2p+1}{p^{2}-p-\ell} \, \binom{p-1}{\ell-(p-1)}}
{\binom{p^{2}-p}{p-1}}\right) 
= 
\frac{1}{\binom{p^{2}-p}{p-1}} \sum^{p-1}_{\ell=1} \left(\frac{\ell}{\ell+p-2} 
\, \binom{p^{2}-2p+1}{\ell} \, \binom{p-1}{\ell} \right).
\]
By Lemma \ref{lem3.1}, we obtain
\begin{align*}
L_{\rm Li}&=\frac{1}{\binom{p^{2}-p}{p-1}} \sum^{p-1}_{\ell=1} \left(\frac{\ell}{\ell + p - 2} \, \binom{p^{2}-2p+1}{\ell} \, \binom{p-1}{\ell}\right) \\
&\geq \frac{1}{\binom{p^{2}-p}{p-1} \, (2p-3)} \sum^{p-1}_{\ell=1} \left(\ell \, \binom{p^{2}-2p+1}{\ell} \, \binom{p-1}{\ell} \right) \\
& > \frac{1}{\binom{p^{2}-p}{p-1} \, (2p-3)} \, (p-3) \,
\binom{p^{2}-p}{p-1} = \frac{p-3}{2p-3}.
\end{align*}
On the other hand,
\begin{align*}
L_{\rm Li} & =\frac{1}{\binom{p^{2}-p}{p-1}} \sum^{p-1}_{\ell=1} \left(\frac{\ell}{\ell+p-2} \binom{p^{2}-2p+1}{\ell} \binom{p-1}{\ell}\right)  \\
 & \leq \frac{1}{\binom{p^{2}-p}{p-1}} \, \frac{p-1}{2p-3} \, \sum^{p-1}_{\ell=1} \binom{p^{2}-2p+1}{\ell} 
 \binom{p-1}{\ell} \\
 & < \frac{1}{\binom{p^{2}-p}{p-1}} \, \frac{p-1}{2p-3} \, \sum^{p-1}_{\ell=0} \binom{p^{2}-2p+1}{\ell} 
 \binom{p-1}{\ell} =\frac{p-1}{2p-3}.
\end{align*}
Thus, $\frac{p-3}{2p-3} < L_{\rm Li} <\frac{p-1}{2p-3}$. The fact that 
\[
\lim_{p \to \infty} \frac{p-3}{2p-3} = \lim_{p \to \infty} \frac{p-1}{2p-3} = \lim_{p \to \infty} L_1 = \frac{1}{2} \mbox{ implies } 
\lim_{p \to \infty} \frac{L_1}{L_{\rm Li}} = 1,
\]
confirming that our cascaded CDC scheme is asymptotically optimal. Figure \ref{Figure40} compares the communication load of our scheme with $L_{\rm Li}$.

\subsection{Concluding Remarks}\label{sec:conclu}

Our constructions highlight the prominent role that combinatorial designs play in minimizing the communication load. We believe that construction routes from known combinatorial structures still have a lot of potential to exploit in improving the performance of distributed computing schemes. Our schemes are \emph{asymptotically} optimal. So are many previously known schemes, most notably the Li-CDC and Jiang-CDC schemes. Unlike those prior schemes, ours have generally improved communication loads, being consistently closer to the theoretical lower bound. Another significant advantage of our schemes lies in the parameters. 

The schemes constructed in Subsection \ref{subsec:scheme1} have $N=Q=v$, with $v$ as specified in Table \ref{table0.1}. This means that the growth of the number $N$ (of input files) and $Q$ (of functions to run) can be nicely calibrated to suit practical constraints. The schemes built in Subsection \ref{subsec:scheme2} have $N=Q=n$, with $n$ being the number of nodes. Since $Q \geq n$, the schemes require only the least possible number of computing nodes and the least number of functions to complete the given task.     

The framework depicted in Figure \ref{fig1} does not appear to have incorporated some error-control mechanism. The assumption is that the whole system is robust, {\it e.g.}, none of the nodes can fail and that the broadcasts are sent and received error-free. In practice, some small number of nodes may fail or a few intermediate values cannot be made available due to transmission errors. The question of error-control coding form CDC schemes seems open for investigation.

\section*{Appendix: Proof of Lemma \ref{lem3.1}}

We begin by establishing the inequality
\begin{equation}\label{for32}
\binom{p^{2}-2p+1}{p-1} > \binom{p-1}{3} \, \binom{p^{2}-2p+1}{p-4}
\end{equation}
by observing directly that 
\begin{align*}
\frac{\binom{p^{2}-2p+1}{p-1}} {\binom{p-1}{3} \, 
\binom{p^{2}-2p+1}{p-4}} & =\frac{6 \, (p^{2}-3p+5) \, (p^{2}-3p+4) \, (p^{2}-3p+3)} {(p-1)^{2} \, (p-2)^{2} \,(p-3)^{2}} \\
&=\frac{6 \, (p^{2}-3p+5) \, (p^{2}-3p+4) \, (p^{2}-3p+3)}{(p^{2}-4p+3)^{2} \,(p^{2}-4p+4)} >1.
\end{align*}
Our next step is to prove the inequality
\begin{equation} \label{for33}
2 \, \binom{p^{2}-2p+1}{p-4} \, \binom{p-1}{ 3} > 
\sum^{p-4}_{\ell=0} (p-3-\ell) \, \binom{p-1}{p-1-\ell} \, \binom{p^{2}-2p+1}{\ell}.
\end{equation}
For any $\ell \in \{0,1,\ldots,p-4\}$, let
\[
d_{\ell} = (p-3-\ell) \, \binom{p-1}{p-1-\ell} \, 
\binom{p^{2}-2p+1}{\ell}.
\]
Hence, as $\ell$ increases in the range $0,1,\ldots,p-5$, the function
\[
\frac{d_{\ell}}{d_{\ell+1}} = \left(\frac{p-3-\ell}{p-4-\ell}\right) \, \frac{(\ell+1)^{2}}{(p^{2}-2p+1-\ell) \, (p-1-\ell)}
\]
is increasing. Hence, for any $\ell \in\{0,1,\ldots,p-5\}$,
\[
\frac{d_{\ell}}{d_{\ell+1}} \leq \frac{d_{p-5}}{d_{p-4}} = \frac{2 \, (p^{2}-8p+16)}{4 \,(p^{2}-3p+6)} < \frac{1}{2},
\]
making it evident that 
\begin{align*}
d_{\ell} &< \frac{1}{2} \, d_{\ell+1} < \ldots < \left(\frac{1}{2}\right)^{p-4-\ell} \, d_{p-4} \mbox{ and}\\
\sum^{p-4}_{\ell=0} d_{\ell} &< \sum^{p-4}_{\ell=0} \left(\frac{1}{2}\right)^{p-4-\ell} \, d_{p-4} = \left(2-\left(\frac{1}{2}\right)^{p-4}\right) \, d_{p-4} < 2d_{p-4},
\end{align*}
settling \eqref{for33}.

Our last step is to establish \eqref{for31}. We use \eqref{for32} and \eqref{for33}, respectively, to get the last two inequalities in the expression
\begin{align*}
& \sum^{p-1}_{\ell=0} \ell \, \binom{p^{2}-2p+1}{\ell} \binom{p-1}{\ell} \\
&= \sum^{p-4}_{\ell=0} \ell \, \binom{p^{2}-2p+1}{\ell} \, 
\binom{p-1}{p-1-\ell} + (p-3) \, \sum^{p-1}_{\ell=p-3} \binom{p^{2}-2p+1}{\ell} \, \binom{p-1}{p-1-\ell} \\
& \ \ + \binom{p-1}{1} \, \binom{p^{2}-2p+1}{p-2} + 2 \, \binom{p^{2}-2p+1}{p-1} \\
&>\sum^{p-4}_{\ell=0} \ell \binom{p^{2}-2p+1}{\ell} \,
\binom{p-1}{p-1-\ell} + (p-3) \, \sum^{p-1}_{\ell=p-3} \binom{p^{2}-2p+1}{\ell} \, \binom{p-1}{p-1-\ell} + 2 \, \binom{p^{2}-2p+1}{p-1} \\
&> \sum^{p-4}_{\ell=0} \ell \, \binom{p^{2}-2p+1}{\ell} \, \binom{p-1}{p-1-\ell} + (p-3) \, \sum^{p-1}_{\ell=p-3} \binom{p^{2}-2p+1}{\ell} \, 
\binom{p-1}{p-1-\ell} + 2 \, \binom{p^{2}-2p+1}{p-4} \, 
\binom{p-1}{3} \\
& > (p-3) \, \sum^{p-4}_{\ell=0} \binom{p^{2}-2p+1}{\ell} \, 
\binom{p-1}{p-1-\ell} + (p-3) \, \sum^{p-1}_{\ell=p-3} \binom{p^{2}-2p+1}{\ell} \, \binom{p-1}{p-1-\ell} = (p-3) \, \binom{p^{2}-p}{p-1}.
\end{align*}
The proof is now complete. \qedsymbol



\end{document}